\documentclass[a4paper, 12pt]{article}
\usepackage{amsmath,amsthm,amssymb}
\usepackage{graphicx}
\usepackage{color}
\usepackage{dsfont}
\usepackage{epstopdf}
\usepackage[hang]{subfigure}
\usepackage{natbib}
\usepackage{ulem}
\usepackage{color}
\usepackage{textcomp}

\newtheorem{thm}{Theorem}
\newtheorem{lem}{Lemma}

\newtheorem{prop}{Proposition}

\newcounter{rmk}

\definecolor{red}{RGB}{139,0,18}
\definecolor{lightred}{RGB}{186,25,31}
\definecolor{blue}{RGB}{0,56,108}
\definecolor{lightblue}{RGB}{69,100,139}
\renewcommand\emph[1]{{\color{red}\itshape #1}}

\def\e{\epsilon}
\def\t{ \mathrm{\scriptscriptstyle T} }

\def\mR{\mathds{R}}

\def\t{\mathrm{\scriptscriptstyle T} }
\def\argmin{\mathop{\arg\min}}

\title{Dimension reduction of high-dimension categorical data with two or multiple responses considering interactions between responses}
\date{}
\author{\small
{Yuehan Yang
}
\\
{\small School of Statistics and Mathematics, Central University of Finance and Economics,} \\
\small{Beijing, 102206, China}
}
\begin{document}
\maketitle
\begin{abstract}
This paper models categorical data with two or multiple responses, focusing on the interactions between responses. We propose an efficient iterative procedure based on sufficient dimension reduction. We study the theoretical guarantees of the proposed method under the two- and multiple-response models, demonstrating the uniqueness of the proposed estimator and with the high probability that the proposed method recovers the oracle least squares estimators. For data analysis, we demonstrate that the proposed method is efficient in the multiple-response model and performs better than some existing methods built in the multiple-response models. We apply this modeling and the proposed method to an adult dataset and right heart catheterization dataset and obtain meaningful results.
\end{abstract}


\section{Introduction}

Categorical data with multiple responses appear in many applications. For example, in a right heart catheterization dataset, some researchers focused on patients who had been diagnosed with sepsis in multiple organs \citep{luo2022reduction}. Meanwhile, patient health data contain records of diagnoses within controlled vocabularies, as well as prescriptions, thus generating categorical features with numbers of levels \citep{tibshirani2021categorical,jensen2012mining}. Models that consider a univariate response are frequently studied. Related research include conducting significance tests \citep{tukey1949comparing,calinski1985clustering} and fusing the levels of categories together in the linear regression setting \citep{breiman2017classification,bondell2009anova,post2013factor,pauger2019bayesian}.

Regarding a univariate response, a valuable technique is to consider an analysis of variance (ANOVA) model that relates the response to categorical predictors. Based on this model, \citet{bondell2009anova} proposed the collapsing and shrinkage in ANOVA (CAS-ANOVA) penalty in which the balanced effects of the categories with certain levels were more prevalent than others. Similarly, \citet{pauger2019bayesian} proposed a Bayesian approach to encourage levels fusion. Further, \citet{tibshirani2021categorical} proposed a SCOPE estimator to fuse the levels of several categories by equivalating the corresponding coefficients. Other studies considered the tree-based model for hierarchical categorical predictors \citep{carrizosa2022tree}, clustering the categories of categorical predictors in generalized linear models \citep{carrizosa2021clustering}, systematic overview of penalty-based methods for categorical data \citep{tutz2016regularized}, etc.

However, in certain applications, the data often comprise two or multiple responses. For example, \citet{little2019statistical} described missing data in which one response is a partially missing variable while the other is the missing indicator of the former. An efficient strategy for studying multiple responses is to introduce sufficient dimension reduction. With two responses, \citet{ding2020double} studied the dimension reduction in survival analysis; therein, one response is of interest while the other is a nuisance variable. Furthermore, \citet{luo2022reduction} studied the dimension reduction regarding the interaction between two responses. \citet{de2011covariate} proposed an iterative two-step procedure to derive a minimal balancing score that connects with the local dimension reduction efficiency in the causal inference.

Other dimension reduction methods for different goals have also been studied in the literature, e.g., missing data analysis \citep{guo2018semiparametric} and causal inference \citep{ma2019robust,luo2020matching}. However, none of the foregoing studies considered categorical data analysis.

Although categorical data draws enormous attention, there is still lack of studies on the multiple-response model, including the efficient algorithm and reliable theoretical guarantees. Thus, the goal of this paper is to construct a model and develop a method for estimating high-dimensional linear models with categorical data considering the interactions between the responses. The first contribution of this paper is to establish the statistical modeling of the categorical data with two or multiple responses. Although many studies have analyzed categorical data with the univariate-response model, it is unclear whether these analyses hold for the data comprising two or multiple responses; moreover, it is also unclear how to study the multiple-response model by techniques used in the univariate response model. To fill in this gap, we establish the dimension reduction theory in categorical data by applying the sufficient dimension reduction technique and locally efficient dimension reduction subspace. We construct a modeling that considers the interactions between response variables, which is a new research problem concern with wide applications in categorical data analysis.

The second contribution of this paper is to propose an efficient iterative procedure for analyzing categorical data via a multiple-response model. The proposed method extends the algorithm of \citet{de2011covariate} for handling multiple responses with categorical covariates. We show that the resulting estimator coincides with the least squares solution in the multiple-response case. Most relevant to our theoretical results, \citet{tibshirani2021categorical} established the theoretical results for their estimator, which fuses the category levels but with the univariate response. Additionally, we show that the proposed procedure is efficient in simulations and applications. We apply the procedure to two real data, adult dataset and right heart catheterization dataset, and both data analyses demonstrate the effectiveness of the model and method.
\section{Models and methods}
\subsection{Notation and models}
We first introduce the notation, as well as the two-response model. Consider an ANOVA model relating two responses and categorical predictors. Set two response variables, $y_1$ and $y_2$, and the categorical predictors, $X$, where $X = (X_1,\dots,X_p) $ and $X_j = (x_{1j},\dots, x_{nj})$. We obtain $x_{ij} \in\{1,\dots,K_j\}$ where $j = 1,\dots,p$. Further, we set the coefficient parameters, $(\mu_1,\theta_1)$ and $(\mu_2,\theta_2)$, corresponding to both responses respectively. Therein, $\mu_1$ and $\mu_2$ are intercepts and $\theta_1, \theta_2 \in \mR^{K_1} \times \cdots \times \mR^{K_p}$ where $\theta_{1j} : = (\theta_{1jk})^{K_j}_{k =1} \in \mR^{K_j}$ and $\theta_{2j} : = (\theta_{2jk})^{K_j}_{k =1} \in \mR^{K_j}$. $\theta_{1jk}$ and $\theta_{2jk}$ are the coefficients of responses $y_1$ and $y_2$ of the $k$th level of the $j$th predictor respectively. Consider the following ANOVA models that relate the two responses and categorical predictors:
\begin{align*}
& y_1 = \mu_1 + \sum_{j = 1}^p \sum^{K_j}_{k = 1}\theta_{1jk} \mathds{1}_{\{x_j = k\}} + \e_1,\\
& y_2 = \mu_2 + \sum_{j = 1}^p \sum^{K_j}_{k = 1}\theta_{2jk} \mathds{1}_{\{x_j = k\}} + \e_2,
\end{align*}
where $\e_1 = (\e_{11},\dots,\e_{1n})$, $\e_2 = (\e_{21},\dots,\e_{2n})$ are the independent zero mean random errors. In this model, the target is to properly estimate $(\mu_1,\theta_1)$ and $(\mu_2,\theta_2)$ regarding the interaction between the two responses.

We introduce the dimension reduction technique into the above model. By setting $R = (R_1,\dots, R_p) \in \mR^{K_1} \times \cdots \times \mR^{K_p}$ such that $R_j:= (\mathds{1}_{x_j = k})^{K_j}_{k =1} \in \mR^{K_j}$, sufficient dimension reduction assumes the existence of a low-dimensional linear combination of $R$. Namely, we obtain $y_1 \perp R| \theta_1 \otimes R$, where $\perp$ denotes the independence between the two responses. Consider the other response, $y_2$, which is interacted with $y_1$. Specifically, for $(y_1,y_2)$, we consider the situation in which they are interacted only through the covariates, i.e., $y_1 \perp y_2|X$. Combined with $y_1 \perp X| \theta_1 \otimes R$, we deduce a brief assumption under sufficient dimension reduction,
\begin{equation}\label{eq assum1}
y_1 \perp y_2| \theta \otimes R.
\end{equation}
Similar to $y_2$, we have $y_2 \perp X| \theta_2 \otimes R$. $\theta_1 \otimes R$ and $\theta_2 \otimes R$ contain the information of $y_1$ and $y_2$ respectively. Particularly, we allow both linear combinations to contain redundant information only from $y_1$ and $y_2$ respectively. The interaction between $y_1$ and $y_2$ is flexible, and this is convenient for the modeling in the latter. We illustrate this theoretically in the next section by showing that assumption~\ref{eq assum1} holds, and a low-dimensional $\theta \otimes R$ exists for sufficient dimension reduction when $\theta_1 \otimes R \perp \theta_2 \otimes R | \theta \otimes R$.

Extending the aforementioned two-response model, we introduce the multiple-response model. For the responses $(y_1,\ldots,y_q)$, we set the coefficient parameters $(\mu_l,\theta_l)$ where $\mu_1,\dots,\mu_q$ are intercepts, and $\theta_1,\dots,\theta_q \in \mR^{K_1} \times \cdots \times \mR^{K_p}$ where $\theta_{lj} : = (\theta_{ljk})^{K_j}_{k =1} \in \mR^{K_j}$. $\theta_{ljk}$ denotes the coefficient to the response, $y_l$, of the $k$th level of the $j$th predictor. Then, we construct the following models for $l = 1,\dots, q$:
\begin{align*}
y_l = \mu_l + \sum_{j = 1}^p \sum^{K_j}_{k = 1}\theta_{ljk} \mathds{1}_{\{x_j = k\}} + \e_l,
\end{align*}
where $\e_l = (\e_{l1},\dots,\e_{ln})$ are the independent zero mean random errors. In this model, the number of estimated parameters depends on the numbers of predictors and responses, and can be much higher than that of the univariate- or two-response model. In this case, sufficient dimension reduction is a useful technique. Similar to the two-response model, we have a brief assumption for the above modeling. Note that $R \in \mR^{K_1} \times \cdots \times \mR^{K_p}$, where $R_j:= (\mathds{1}_{x_j = k})^{K_j}_{k =1} \in \mR^{K_j}$. For each response, the linear combination of $R$ preserves all the information in $R$ of modeling the response, i.e., for $l = 1,\dots,q$,
\[ y_l \perp R | \theta_l \otimes R.\]
We assume that the responses interacted only through the covariates. Thus, this assumption is deduced by combining it with the above model, as follows:
\begin{equation}\label{eq assum2}
y_1 \perp \cdots \perp y_q| \theta \otimes R.
\end{equation}
We allow each linear combination to carry redundant information from its response only. $\theta \otimes R$ differs from $\theta_l \otimes R$ for $l = 1,\dots,q$, where the former only preserves the information of $R$ regarding the interactions between the responses and does not need to preserve the information about the modeling of each response, such as the latter. Based on this assumption, the dimension reduction technique is allowed for the loss of information about the modeling of each response. Additionally, we prove in the following that $\theta \otimes R$ exists and that assumption~\ref{eq assum2} holds when $\theta_{l_1} \otimes R \perp \theta_2 \otimes R  |\theta \otimes R$.

\subsection{Methods}

To estimate $(\mu_1,\theta_1)$ and $(\mu_2,\theta_2)$ based on the two-response model under the sufficient dimension reduction assumption, we consider an iterative two-step procedure for the two-response model. In the first step, consider the first response $y_1$ and categorical predictors $X$. We find a subset of $X$, namely $X_{\mathcal{A}^{[1]}}$ such that $\mathcal{A}^{[1]} \subset \{1,\dots,p\}$ and $y_1 \perp X | X_{\mathcal{A}^{[1]}}$. This is directly followed from $y_1 \perp | \theta_1 \otimes R$ where $R_j:= (\mathds{1}_{x_j = k})^{K_j}_{k =1} \in \mR^{K_j}$. In the second step, based on the first step, we build the model with $y_2$ and $X_\mathcal{A}^{[1]}$ and obtain the subset $\mathcal{A}^{[2]}$ of $\mathcal{A}^{[1]}$ that $\mathcal{A}^{[2]} \subset \mathcal{A}^{[1]}$. Then, we repeat the above two steps and update the set $\mathcal{A}^{[k]}$ until convergence.

The above method is inspired by \citet{de2011covariate} and \citet{yyh202011}. The former proposed an iterative two-step procedure for sufficient dimension reduction to derive a minimum balancing score, and the latter introduced an iterative algorithm for the penalized regularization of the linear regression model. The iteration algorithm exhibited two advantages: 1) in each step, the model deals with the reduced feature space obtained from the previous step, rendering the computational complexity comparable with that of low-dimensional models; 2) in each iteration, the challenge of estimating a two-response model is reduced to that of estimating a univariate-response model.

Consider the categorical predictors with a univariate response. Early work includes greedy approaches, such as the classification and regression tree (CART) algorithm \citep{breiman2017classification} and penalized methods, e.g., CAS-ANOVA \citep{bondell2009anova} and SCOPE \citep{tibshirani2021categorical} in which CAS-ANOVA applies the penalty that can be seen as an `all-pairs' version of the fused Lasso:
\[ \sum^p_{j=1} \sum^{K_j}_{k=2} \sum^{k-1}_{l = 1} w_{j,kl} |\theta_{jk} - \theta_{jl}|, \]
where $w_{j,kl}$ are the weights for balancing the effects of certain levels of categories that are more prevalent than others. As shown by \citet{tibshirani2021categorical}, a drawback of CAS-ANOVA is that it might not perform well for groups of unequal sizes. Based on that, SCOPE uses the following penalty:
\[\sum^p_{j=1} \sum^{K_j-1}_{k=1} \rho_j (\theta_{j(k+1)} - \theta_{j(k)}), \]
where $\rho_j$ represents concave non-decreasing penalty functions, and $\theta_{j(1)} \leqslant \cdots \leqslant \theta_{j(K_j)}$ are the order statistics of $\theta_j$. Inspired by the above work, the iterative two-step procedure for the two-response model is introduced in the following step. First, we discuss the estimation of the intercepts, $\mu_1$ and $\mu_2$. To ensure identifiability, we set $\theta \in \Theta$ where $\Theta = \Theta_1 \times \cdots \times \Theta_p$ and
\[\Theta_j = \{ \theta_j \in \mR^{K_j}: \sum^{K_j}_{k = 1} \sum^n_{i = 1}  \mathds{1}_{\{x_{ij} = k\}} \theta_{jk}= 0 \}.\]
Thus, the estimated intercepts of $\mu_1$ and $\mu_2$ can be calculated as $\hat \mu_1 = \sum^n_{i=1} y_{1i}/n$ and $\hat \mu_2 = \sum^n_{i=1} y_{2i}/n$, respectively. Recall that $\theta_{1j} : = (\theta_{1jk})^{K_j}_{k =1} \in \mR^{K_j}$ and $\theta_{2j} : = (\theta_{2jk})^{K_j}_{k = 1} \in \mR^{K_j}$. Set $\mathcal{A} \subset \{1,\dots,p\}$ as an index set and denote $\theta_{1\mathcal{A}} := (\theta_{1j}, j \in \mathcal{A})$ and $\theta_{2\mathcal{A}}:= (\theta_{2j}, j \in \mathcal{A})$. Then we obtain the following algorithm:
\begin{center}
\textit{Iterative two-step algorithm}
\end{center}
\begin{itemize}
\item[1.] Start with the universal sets $\mathcal{A}_1 = \mathcal{A}_2 = \{1,\dots, p\}$.
\item[2.] For $y_1$, solve the following function with $\hat \mu_1 = \sum^n_{i=1} y_{1i}/n$ and update the nonzero index set $\mathcal{A}_1$ of $\hat \theta_1$:
\begin{equation}\label{eq solution 1}
\hat \theta_1 =\argmin_{\hat \theta_{1\mathcal{A}_2^c} = 0} \Big\{ \dfrac{1}{2n} \Big\| y_1 - \hat \mu_1 -  \sum_{j = 1}^p \sum^{K_j}_{k = 1}\theta_{1jk} \mathds{1}_{\{x_j = k\}} \Big\|^2_2 + \sum^p_{j=1} \sum^{K_j-1}_{k=1} \rho_{1j} (\theta_{1j(k+1)} - \theta_{1j(k)})\Big\}.
\end{equation}
\item[3.] For $y_2$, solve the following function with $\hat \mu_2 = \sum^n_{i=1} y_{2i}/n$, followed by updating the nonzero index set $ \mathcal{A}_2$ of $\hat \theta_2$:
\begin{equation}\label{eq solution 2}
\hat \theta_2 = \argmin_{\hat \theta_{2\mathcal{A}_1^c} = 0} \Big\{ \dfrac{1}{2m} \Big\| y_2 - \hat \mu_2-  \sum_{j = 1}^p \sum^{K_j}_{k = 1} \theta_{2jk} \mathds{1}_{\{x_j = k\}} \Big|^2 + \sum^p_{j=1} \sum^{K_j-1}_{k=1} \rho_{2j} (\theta_{2j(k+1)} - \theta_{2j(k)})\Big\}.
\end{equation}
\item[3.] Repeat 2 - 3 until convergence.
\end{itemize}
In the above algorithm, $\rho_{1j}$ and $\rho_{2j}$ are the same type of penalty functions. In the following, we use the minimax concave penalty (MCP) \citep{zhang2010mcp} followed by SCORE \citep{tibshirani2021categorical}. The index sets for modeling $y_1$ and $y_2$ are at convergence respectively in this algorithm. For instance, they are nested during the iterations. $\mathcal{A}_1$, when $\theta_{1jk} = 0$ for all $k = 1,\dots, K_j$, we delete $j$ from the index set.

Following the same setting as in the two-response model, we introduce the following iterative procedure:
\begin{center}
\textit{Iterative $q$-step algorithm}
\end{center}
\begin{itemize}
\item[1.] Start with the universal set $\mathcal{A}_0 = \{1,\dots, p\}$.
\item[2.] For each $l = 1,\dots,q$, solve the following function with $\hat \mu_l = \sum^n_{i=1} y_{li}/n$ and update the nonzero index set $\mathcal{A}_l$ of $\hat \theta_l$:
\begin{equation}\label{eq solution}
\hat \theta_l = \argmin_{\hat \theta_{l,\mathcal A_{l-1}^c} = 0} \Big\{\dfrac{1}{2n}\Big\| y_l - \hat \mu_l-  \sum_{j = 1}^p \sum^{K_j}_{k = 1} \theta_{ljk} \mathds{1}_{\{x_j = k\}} \Big\|^2_2 + \sum^p_{j=1} \sum^{K_j-1}_{k=1} \rho_{lj} (\theta_{lj(k+1)} - \theta_{lj(k)})\Big\}.
\end{equation}
\item[3.] Repeat 2 until convergence.
\end{itemize}

Based on the theoretical results and with suitable tuning parameters, the iteration converges to the oracle least squares solution in the first two steps and the first $q$ steps for the two-response model and the multiple-response model, respectively. Although the tuning parameters cannot be accurately selected in simulations and applications, the iteration still converges in the first two or $q$ steps in most simulations according to our numerical experience. This advantage greatly reduces the computational cost of the proposed method. We thus introduce the following $q$ steps method:
Set the universal set $\mathcal{A}_0 = \{1,\dots, p\}$. For each $l = 1,\dots,q$, solve the following function with $\hat \mu_l = \sum^n_{i=1} y_{li}/n$ and obtain the nonzero index set $\mathcal{A}_l$ of $\hat \theta_l$:
\begin{equation*}
\hat \theta_l = \argmin_{\hat \theta_{l,\mathcal A_{l-1}^c} = 0} \Big\{\dfrac{1}{2n}\Big\| y_l - \hat \mu_l-  \sum_{j = 1}^p \sum^{K_j}_{k = 1} \theta_{ljk} \mathds{1}_{\{x_j = k\}} \Big\|^2_2 + \sum^p_{j=1} \sum^{K_j-1}_{k=1} \rho_{lj} (\theta_{lj(k+1)} - \theta_{lj(k)})\Big\}.
\end{equation*}

\section{Theoretical results}

\subsection{Two-response model}
For the two-response model, we assume that $y_1 \perp y_2| \theta \otimes R$ based on the sufficient dimension reduction. Further, we require that the independence between $y_1$ and $y_2$ induces the independence between the directions of $R$ that are associated with these two response variables. This is a regular requirement supported by $y_1 \perp y_2| \theta \otimes R$ since the latter assumption constrains the additional association between $y_1$ and $y_2$. Both requirements are used to introduce the following proposition.

We first introduce the uniqueness of the estimator. Usually, the sufficient dimension reduction technique may assume that $X$ follows a multivariate normal distribution. However, $X$ is the categorical predictor matrix in this paper. Based on this setting, we specifically consider that the locally efficient dimension reduction subspace for $y_1|X$ and $y_2|X$ can be spanned by some $(\alpha_1, \alpha_2)$ and $(\alpha_2, \alpha_3)$, respectively, since $\alpha_1$, $\alpha_2$, and $\alpha_3$ are orthogonal to one another. This requirement can be explained by the following result:
\begin{prop}\label{prop 1}
For the two-response model, the globally efficient dimension reduction subspace exists if and only if the intersection of the locally efficient dimension reduction subspaces satisfies the following:
\begin{equation}\label{eq assum3}
\theta_1 \otimes R \perp \theta_2 \otimes R  |\theta \otimes R.
\end{equation}
In the meantime, the minimizers of \eqref{eq solution 1} and \eqref{eq solution 2} are unique in both responses.
\end{prop}

This result is mainly based on Proposition 3 of \citet{tibshirani2021categorical} and Theorem 3 of \citet{luo2022reduction}. Thus, we omit the proof here. \eqref{eq assum3} indicates that conditional on the common effect of $X$ on $y_1$ and $y_2$. The effect of $X$ on $y_1$ is independent of the effect of $X$ on $y_2$. Further, we establish the requirements for ensuring the consistency of the proposed algorithm. In this section, we follow the framework and notation of \citet{tibshirani2021categorical}. We consider the high-dimensional setting in which $\sum_{j = 1}^{p} K_j$ can be much larger than the sample size. In the meantime, we set
\[s_{1j} := |\{\theta_{1j1},\dots,\theta_{1jK_j}\}| ~\text{and}~ s_{2j} := |\{\theta_{2j1},\dots,\theta_{2jK_j}\}|,  \]
where $\sum_j^p (s_{1j} -1)$ and $\sum_j^p (s_{2j} -1)< n$. These two index sets indicate the clusters where the coefficients corresponding to a given predictor belong. It is natural to assume that they are lower than the sample size, as well as ensure the following functions have unique solutions. We then define the following two oracle least squares estimates:
\begin{align}\label{eq ols}
& \hat \theta^0_1 :=\argmin_{\theta \in \Theta_0} \dfrac{1}{2n}  \Big\| y_1 - \hat \mu - \sum^p_{j = 1} \sum^{K_j}_{k = 1} \theta_{jk} \mathds{1}_{\{ x_j = k \}} \Big\|^2_2 ~\text{and} \nonumber\\
& \hat \theta^0_2 :=\argmin_{\theta \in \Theta_0} \dfrac{1}{2n}  \Big\| y_2 - \hat \mu - \sum^p_{j = 1} \sum^{K_j}_{k = 1} \theta_{jk} \mathds{1}_{\{ x_j = k \}} \Big\|^2_2,
\end{align}
where $\Theta_0 = \{ \theta^0 \in \Theta: \theta^0_{ljk_1} = \theta^0_{ljk_2} ~\text{whenever}~ \theta_{ljk_1} = \theta_{ljk_2}, l = 1,2\}$, that is, we assume that the fused signals are the oracular knowledge in both estimates. In this case, we set $\hat \theta^0_1 = A_1 y_1$ and $\theta^0_2 = A_2y_2$ with fixed matrixes, $A_1$ and $A_2$, respectively. For the proposed algorithm with penalties $\rho_{1j}$ and $\rho_{2j}$, we consider the MCP penalty \citep{zhang2010mcp} of both functions of the iterative two-step algorithm for simplicity of the proof, as follows:
\begin{align*}
& \rho_{1j} =  \int^x_0 \lambda_{1j}\Big(1 - \dfrac{t}{\gamma_{1j}\lambda_{1j}}\Big)_{+} dt, \\
& \rho_{2j} = \int^x_0 \lambda_{2j}\Big(1 - \dfrac{t}{\gamma_{2j}\lambda_{2j}}\Big)_{+} dt.
\end{align*}
For the signals that differ in each $j = 1,\dots,p$ and $l = 1,2$, we set
\[ \Delta(\theta_{lj}) := \min_{k_1,k_2} \{ |\theta_{ljk_1} - \beta_{ljk_2}|: \theta_{ljk_1} \neq \theta_{ljk_2} \}.  \]
Further, we introduce the following three notations, for $n_{jk} = \|\mathds{1}_{x_{j} = k}\|_1$,
\[ m_{j, \min} = \min_k n_{jk} , \]
and for the fused signals in each $j$,
\[ n_{j,\min} = \min_{l = 1,2}\min_{k_1} \sum_{k_2: \theta_{ljk_2} = \theta_{ljk_1}} n_{ljk_2}, n_{j,\max} = \max_{l = 1,2}\max_{k_1} \sum_{k_2: \theta_{ljk_2} = \theta_{ljk_1}} n_{ljk_2}.\]
Set $c_{\min}: = \min( (\max_r (A_1 A^\t_1)_{rr})^{-1}, (\max_l (A_2 A^\t_2)_{rr})^{-1})$, $\lambda_j  = \min(\lambda_{1j},\lambda_{2j})$. We then introduce the following result:
\begin{thm}\label{thm 1}
Suppose that for $j = 1,\dots,p$, we have $\eta/\max(s_{1j},s_{2j}) \leqslant n_{j,\min}/n \leqslant n_{j,\max}/n \leqslant 1/\eta \min(s_{1j},s_{2j}) $ where $\eta \in (0,1]$. Assume $$\Delta(\theta_{lj}) \geqslant (4+ 3\sqrt{2}/\eta) \sqrt{\max(\gamma_{1j},\gamma_{2j}) \gamma^*_j}\max(\lambda_{1j},\lambda_{2j}), $$ where $\gamma^*_j = \max\{\gamma_{1j}, \gamma_{2j}, \eta s_{1j}, \eta s_{2j}\}$. Then with probability at least $1 - 4 \exp \big( - (n_{j,\min} \wedge c_{\min} ) \eta \gamma^*_j s_{1j} \lambda^2_{1j}/(8 \sigma^2) + \log(2K_j) \big)$, the proposed method converges in the first two steps and $$\hat \theta_1  = \hat \theta^0_1 ~\text{and}~\hat \theta_2 = \hat \theta^0_2.$$
\end{thm}

The above result is mainly based on Theorems 5 and 6 of \citet{tibshirani2021categorical}. We generalize their results in the two-response model.

\subsection{Multiple-response model}
In the previous section, we consider the two-response model, i.e., $q = 2$. Here, we consider the multiple responses, i.e., $q > 2$, and the number of responses, in this case, can also be high-dimensional. The theoretical results of the multiple-response model are similar to that of the two-response model. For the uniqueness of the estimator under the multiple-response model, where $l = 1,\dots,q$, following Proposition~\ref{prop 1}, we easily obtain a similar result such that for each response, the minimizer of \eqref{eq solution} is unique. Further, the globally efficient dimension exists if and only if the intersection between each pair of the local efficient dimension reduction subspaces, denoted as $l_1, l_2 = 1,\dots,q$, satisfies
\[ \theta_{l_1} \otimes R \perp \theta_{l_2} \otimes R  |\theta \otimes R.\]
For $l = 1,\dots, q$, set $s_{lj} := |\{\theta_{1j1},\dots,\theta_{1jK_j}\}|$, $s_{\min,j} = \min(s_{1j},\dots,s_{qj})$, and $s_{\max,j} = \max(s_{1j},\dots,s_{qj})$. We assume $\sum_j^p (s_{\max,j} -1) < n$ to ensure that the following function has unique solution for each $l$,
\[\hat \theta^0_l :=\argmin_{\theta_l \in \Theta_0} \dfrac{1}{2n}  \Big\| y_l - \hat \mu - \sum^p_{j = 1} \sum^{K_j}_{k = 1} \theta_{ljk} \mathds{1}_{\{ x_j = k \}} \Big\|^2_2.\]
Similar to the previous section, we set notations of $\Theta_0$, $\rho_{lj}$, $\Delta(\theta_{lj})$, $n_{jk}$, $m_{j,\min}$, $n_{j,\min}$, $n_{j,\max}$, and $c_{\min}$, and further set $\lambda_{\max,j} = \max(\lambda_{1j},\dots,\lambda_{qj})$ and $\gamma_{\max,j} = \max(\gamma_{1j},\dots,\gamma_{qj})$. We obtain the following result for the multiple-response model.
\begin{thm}\label{thm 2}
Suppose that for $j = 1,\dots,p$, we have $\eta/s_{\max,j} \leqslant n_{j,\min}/n \leqslant n_{j,\max}/n \leqslant 1/\eta s_{\min,j} $ where $\eta \in (0,1]$.
Assume
$$\Delta(\theta_{lj}) \geqslant (4+ 3\sqrt{2}/\eta) ( \gamma_{\max,j} \gamma^*_j)^{1/2} \lambda_{\max,j}, $$
where $\gamma^*_j = \max( \gamma_{\max,j} \eta s_{\max,j})$.
Then a with probability at least $1 - 4 \exp \big( - (n_{j,\min} \wedge c_{\min} ) \eta \gamma^*_j s_{1j} \lambda^2_{1j}/(8 \sigma^2) + \log(q \times K_j) \big)$, the proposed method converges in the first $q$ steps and for $l = 1,\dots, q$ that $$\hat \theta_l  = \hat \theta^0_l.$$
\end{thm}

\section{Simulations}

In this section, we demonstrate the performance of the proposed method on simulated data. As we are aware, no existing methods provide estimation under this model with two or multiple responses. Thus, we illustrate the effectiveness of the proposed method in two parts. In the first part, we show the performance of the proposed method in the two-response model compared with the performance in the univariate response model. The latter can be seen as the original SCOPE proposed by \citep{tibshirani2021categorical}. In the second part, we focus on the two-response model and compare the proposed method with several existing methods, i.e., lasso \citep{tibshirani1996lasso} and random forest \citep{breiman2001random}. The former applies the R package glmnet \citep{friedman2010regularization} and the latter applies the R package randomForest \citep{liaw2002classification}. Since the lasso is unsuitable for the model, we follow the setting from \citet{tibshirani2021categorical} and construct dummy variables for all the categories for the lasso. Each simulation is repeated 100 times.

In the simulations, we consider the two-response model comprising $n = 200$, $p = 100$, and the following models for the simulation studies:
\begin{align*}
& y_1 = \mu_1 + \sum_{j = 1}^p \sum^{K_j}_{k = 1}\theta_{1jk} \mathds{1}_{\{x_j = k\}} + \e_1,\\
& y_2 = \mu_2 + \sum_{j = 1}^p \sum^{K_j}_{k = 1}\theta_{2jk} \mathds{1}_{\{x_j = k\}} + \e_2,
\end{align*}
where the data are simulated following \citet{tibshirani2021categorical}. Namely, the errors are independently distributed as $\mathcal{N}(0,\sigma^2)$ with $\sigma = 1$ and $2.5$. To construct $x_{ij}$, we obtain $W_{ij}$ from $\mathcal{N}_p(0,\Sigma)$; we set the diagonal elements of $\Sigma$ equal to 1 and the off-diagonal elements of $\Sigma$ are selected such that $u_{ij} = \Phi^{-1}(W_{ij})$ with $\text{corr}(u_{ij}, u_{ik}) = \rho$ for $j \neq k$. Thereafter, we obtain $x_{ij} = [24u_{ij}]$ for the number of levels $K_j=24$. We consider the following two scenarios both having two responses with two vectors of true coefficients:

\noindent \textbf{Scenario 1: Sparse case}
\begin{align*}
&\theta_{j1} = (\overbrace{-3,\dots,-3}^{\text{10 times}},\overbrace{0,\dots,0}^{\text{4 times}},\overbrace{3,\dots,3}^{\text{10 times}}) \ \text{for} \ j = 1,2,3, \ \text{and} \ \theta_{j1} = 0 \ \text{otherwise}\\
&\theta_{j2} = (\overbrace{-3,\dots,-3}^{\text{8 times}},\overbrace{0,\dots,0}^{\text{8 times}},\overbrace{3,\dots,3}^{\text{8 times}}) \ \text{for} \ j = 1,2,3, \ \text{and} \ \theta_{j2} = 0 \ \text{otherwise}.
\end{align*}

\noindent \textbf{Scenario 2: Less sparse case}
\begin{align*}
&\theta_{j1} = (\overbrace{-2,\dots,-2}^{\text{16 times}},\overbrace{3,\dots,3}^{\text{8 times}}) \ \text{for} \ j = 1,\dots,25, \ \text{and} \ \theta_{j1} = 0 \ \text{otherwise}\\
&\theta_{j2} = (\overbrace{-3,\dots,-3}^{\text{12 times}},\overbrace{3,\dots,3}^{\text{12 times}}) \ \text{for} \ j = 1,\dots,10, \ \text{and} \ \theta_{j2} = 0 \ \text{otherwise}.
\end{align*}

First, we demonstrate the performance of the proposed method dealing with the two-response model compared with the performance in the univariate-response model. Two measures are used, namely, $l_2$-norm error and mean square error (MSE). The results of both scenarios are presented in Tables~\ref{table 1} and \ref{table 2}. As presented in Table~\ref{table 1} under Scenario 1, for the situation involving the two-response model, the methods proposed for the univariate-response model are no longer suitable. Both $l_2$-norm and MSE are significantly reduced when we consider the two-response model instead of the univariate-response model. For example, when we consider the data setting comprising $\sigma = 1$ and $\rho = 0 $, the two-response model can reduce the $l_2$-norm and MSE of the univariate-response model by 77\% - 97\%. The results under Scenario 2 are presented in Table~\ref{table 2}. In Scenario 2, we consider the less sparse case, while the comparison results are similar to those under Scenario 1.

\begin{table}[htb]
\centering
\small
\caption{Performance of the proposed method under Scenario 1. The original version indicates that we treat the data as the univariate-response model. Response 1 and Response 2 are the results indicating that we treat the data as the two-response model  \label{table 1}}
\begin{tabular}{ccccccc}
\hline
&   & $l_2$-norm      & MSE     &   & $l_2$-norm      & MSE     \\  \hline
Original version  & $\sigma = 1$  & 7.644       & 2.773       & $\sigma = 2.5$  & 10.267      & 4.366       \\
&   & (3.099) & (1.535) &   & (1.011) & (1.128) \\
Response 1  & $\rho = 0$  & 1.810       & 0.057       & $\rho = 0$  & 9.086       & 3.497       \\
&   & (0.579) & (0.114) &   & (1.711) & (1.415) \\
Response 2  &   & 1.769       & 0.077       &   & 9.498       & 3.923       \\
&   & (0.720) & (0.174) &   & (2.019) & (1.758) \\  \hline
Original version  & $\sigma = 1$  & 10.119      & 3.632       & $\sigma = 2.5$  & 14.734      & 6.859       \\
&   & (1.693) & (1.104) &   & (4.924) & (3.385) \\
Response 1  & $\rho = 0.8$  & 3.054       & 0.516       & $\rho = 0.8$  & 12.473      & 6.645       \\
&   & (3.020) & (1.263) &   & (2.920) & (3.560) \\
Response 2  &   & 3.154       & 0.609       &   & 13.551      & 7.023       \\
&   & (3.455) & (1.721) &   & (3.452) & (3.472) \\  \hline
\end{tabular}
\end{table}

\begin{table}[htb]
\centering
\caption{Performance of the proposed method under Scenario 2. The original version indicates that we treat the data as the univariate-response model. Response 1 and Response 2 are the results indicating that we treat the data as the two-response model  \label{table 2}}
\begin{tabular}{ccccccc}
\hline
&   & $l_2$-norm      & MSE     &   & $l_2$-norm      & MSE     \\  \hline
Original version  & $\sigma = 1$  & 5.710       & 0.039       & $\sigma = 2.5$  & 8.783       & 4.008       \\
&   & (0.578) & (0.017) &   & (6.814) & (10.576)  \\
Response 1  & rho = 0 & 5.710       & 0.039       & rho = 0 & 9.115       & 3.464       \\
&   & (0.578) & (0.017) &   & (4.433) & (6.363) \\
Response 2  &   & 5.559       & 0.038       &   & 8.449       & 2.555       \\
&   & (0.583) & (0.017) &   & (3.150) & (4.587) \\  \hline

Original version  & $\sigma = 1$  & 16.003      & 9.985       & $\sigma = 2.5$  & 34.081      & 26.302      \\
&   & (0.578) & (0.017) &   & (12.081)  & (16.623)  \\
Response 1  & $\rho = 0.8$  & 10.643      & 6.565       & $\rho = 0.8$  & 21.502      & 18.706      \\
&   & (0.578) & (0.017) &   & (8.724) & (12.488)  \\
Response 2  &   & 10.850      & 6.874       &   & 19.846      & 17.232      \\
&   & (0.583) & (0.017) &   & (8.557) & (12.754)  \\  \hline
\end{tabular}
\end{table}

\begin{figure}[!htp]
\centering
\subfigure[Response 1, $\sigma = 1$, and $\rho = 0$]
{\includegraphics[width=.45\textwidth,height=.25\columnwidth]{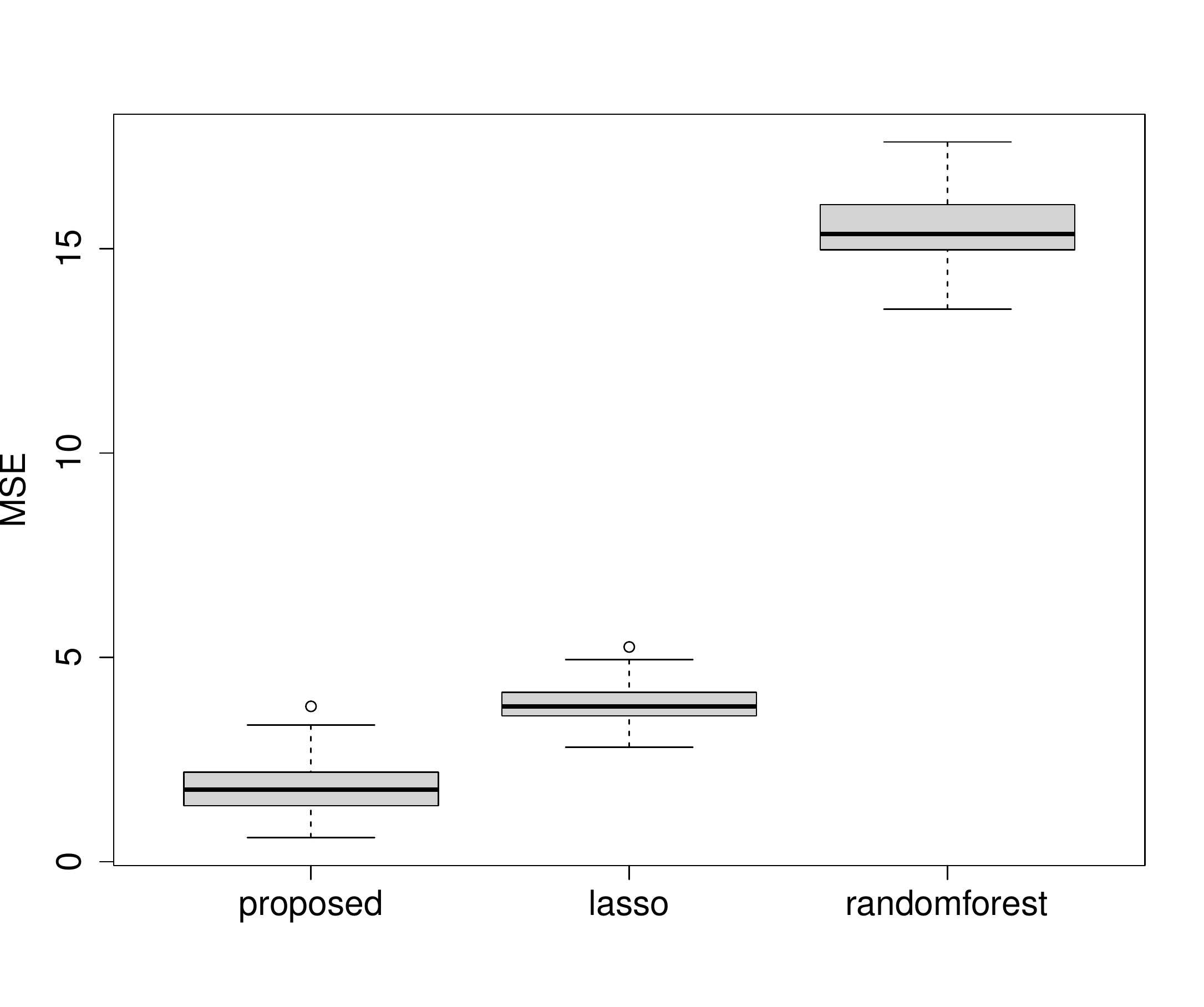}}
\subfigure[Response 1, $\sigma = 1$, and $\rho = 0.8$]
{\includegraphics[width=.45\textwidth,height=.25\columnwidth]{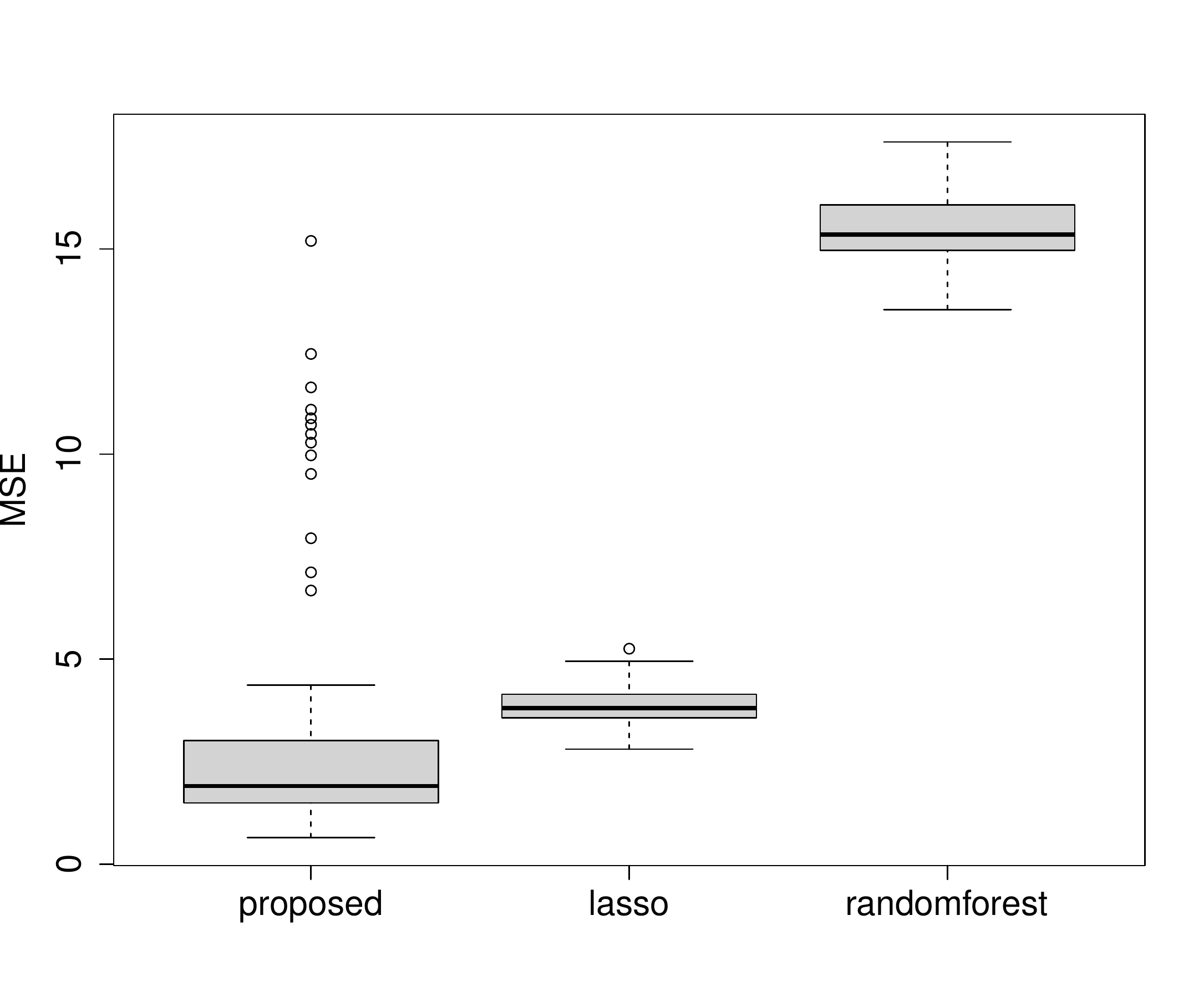}}
\subfigure[Response 1, $\sigma = 2.5$, and $\rho = 0$]
{\includegraphics[width=.45\textwidth,height=.25\columnwidth]{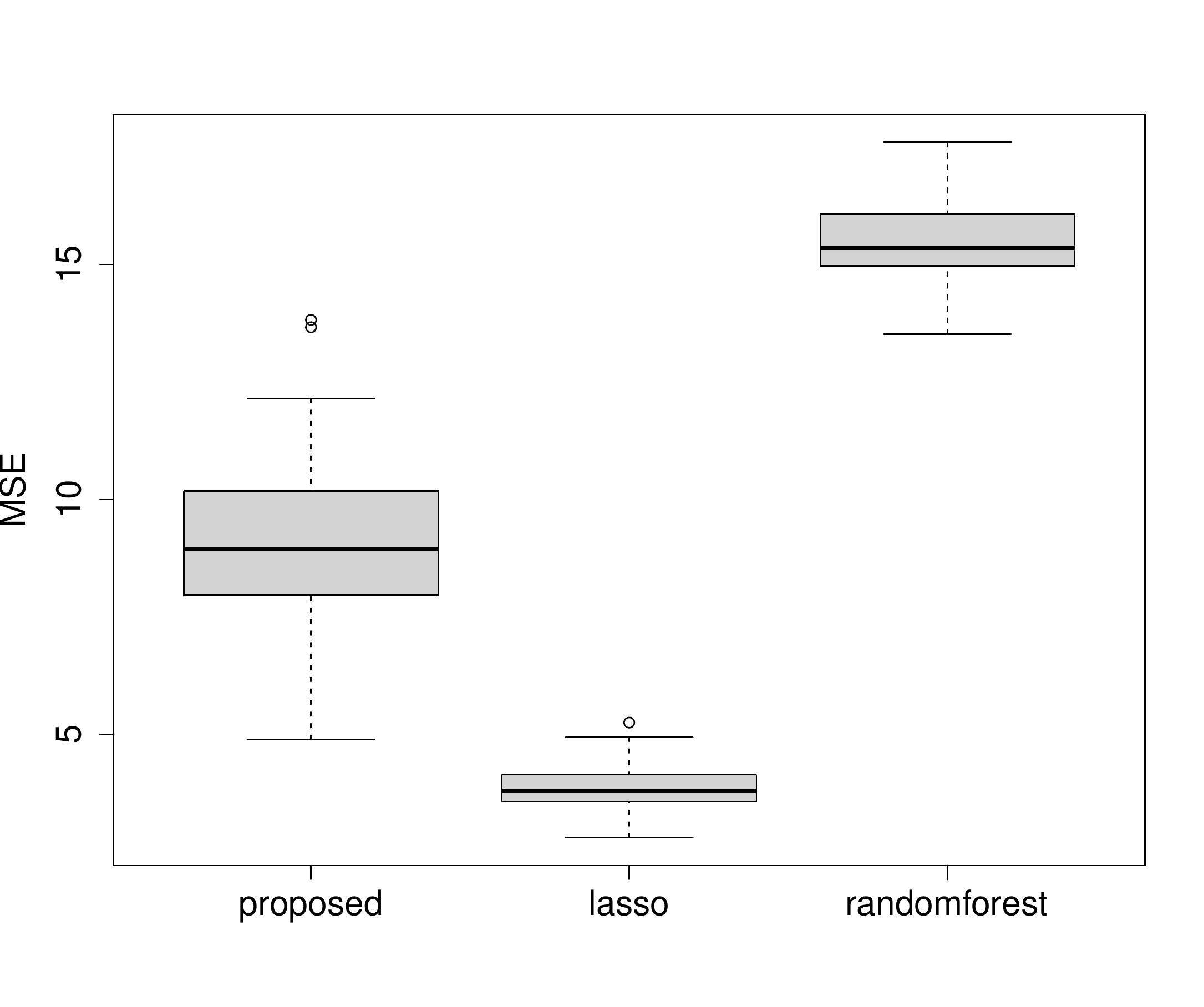}}
\subfigure[Response 1, $\sigma = 2.5$, and $\rho = 0.8$]
{\includegraphics[width=.45\textwidth,height=.25\columnwidth]{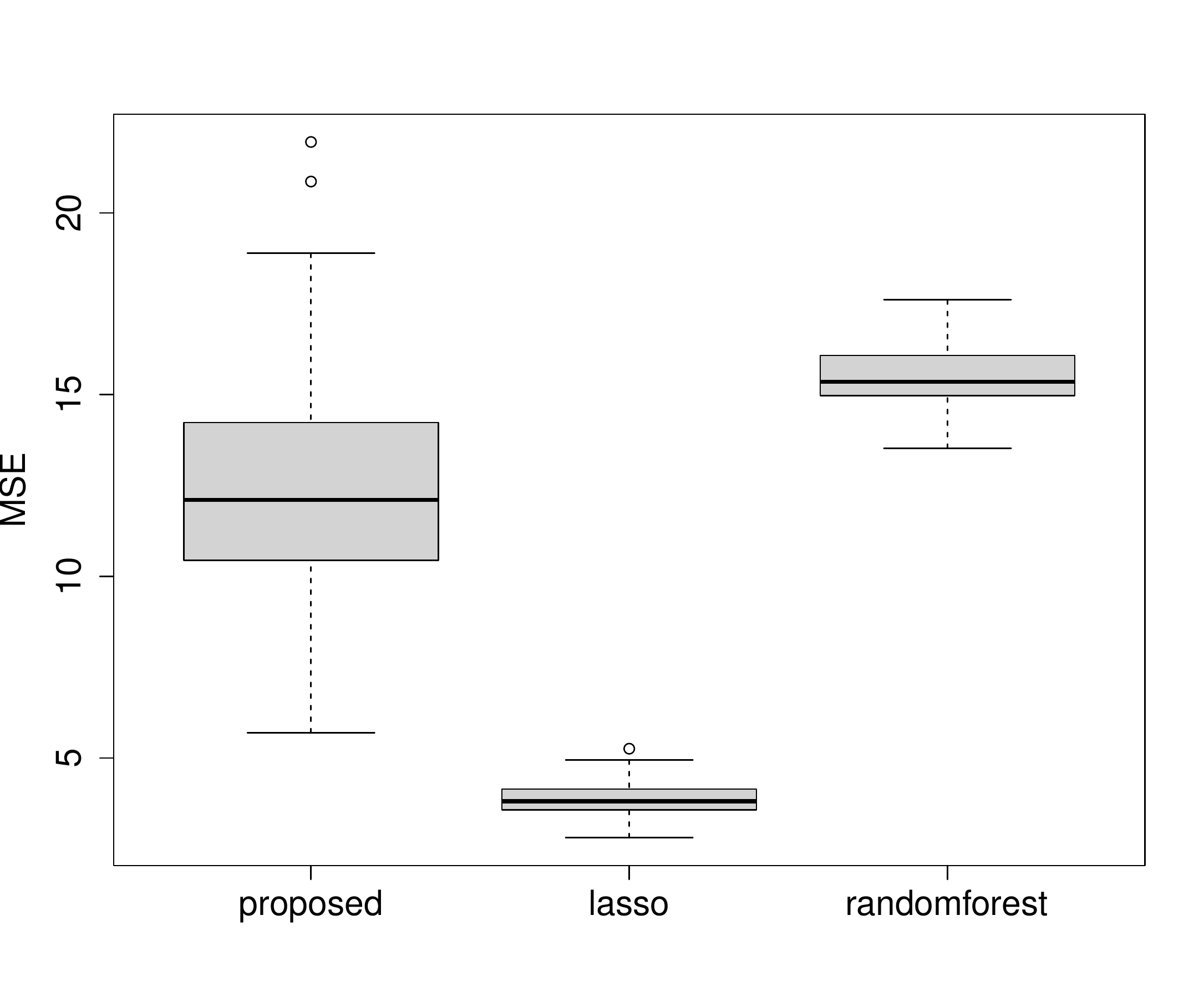}}
\subfigure[Response 2, $\sigma = 1$ and $\rho = 0$]
{\includegraphics[width=.45\textwidth,height=.25\columnwidth]{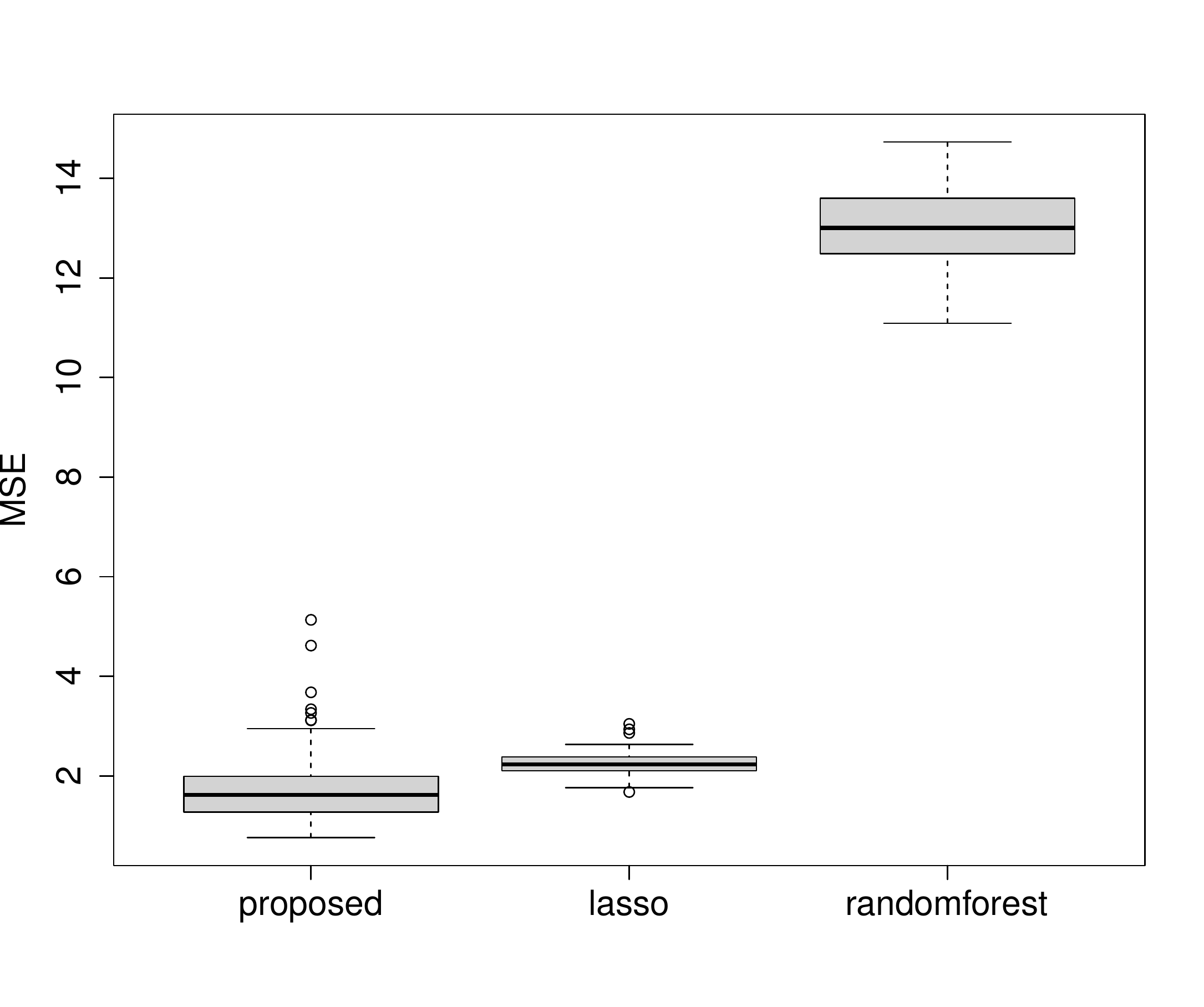}}
\subfigure[Response 2, $\sigma = 1$, and $\rho = 0.8$]
{\includegraphics[width=.45\textwidth,height=.25\columnwidth]{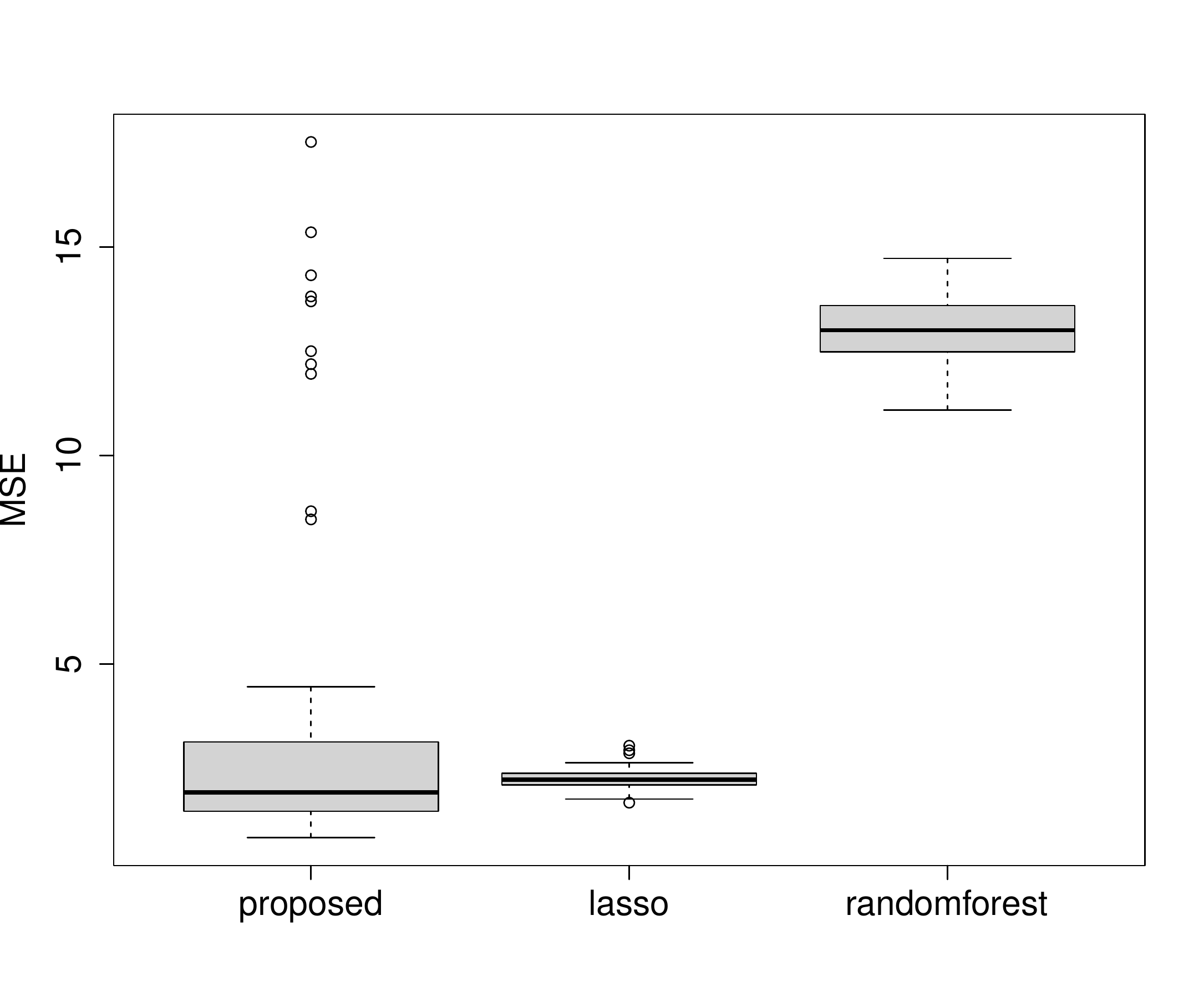}}
\subfigure[Response 2, $\sigma = 2.5$, and $\rho = 0$]
{\includegraphics[width=.45\textwidth,height=.25\columnwidth]{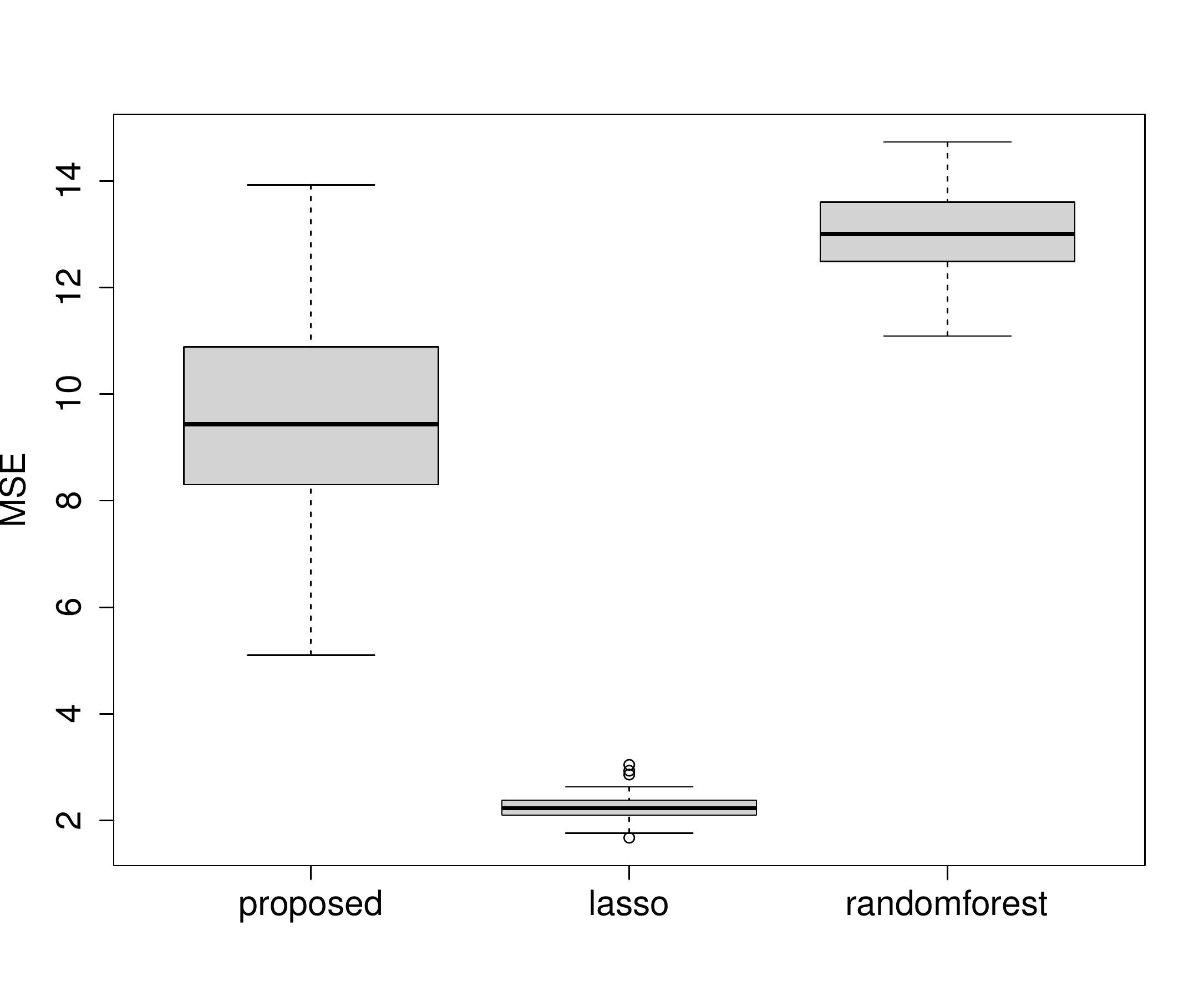}}
\subfigure[Response 2, $\sigma = 2.5$, and $\rho = 0.8$]
{\includegraphics[width=.45\textwidth,height=.25\columnwidth]{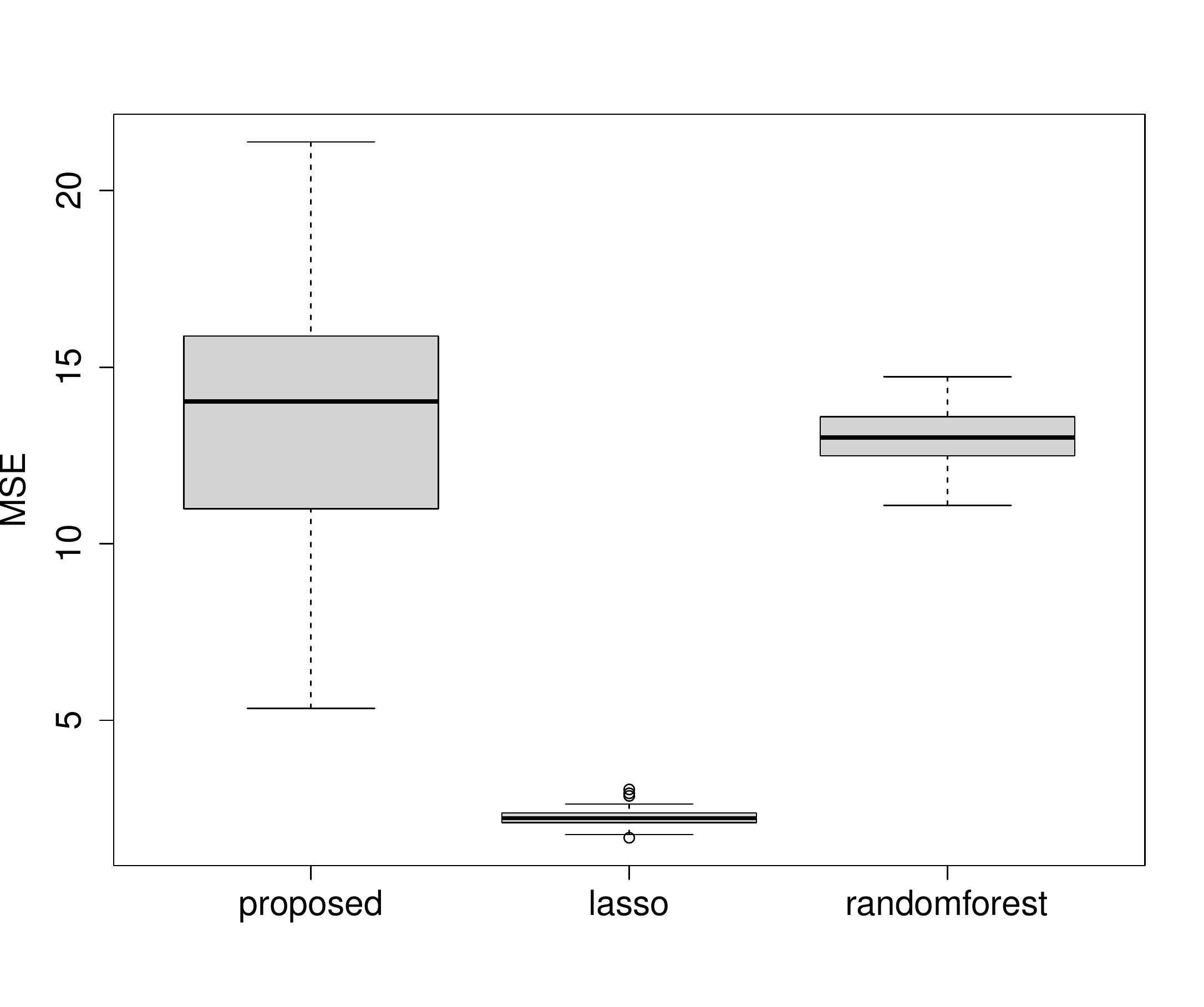}}
\caption{Comparisons of the performances of the proposed method, lasso, and random forest under Scenario 1}
\label{fig 1}
\end{figure}

\begin{figure}[!htp]
\centering
\subfigure[Response 1, $\sigma = 1$, and $\rho = 0$]
{\includegraphics[width=.45\textwidth,height=.25\columnwidth]{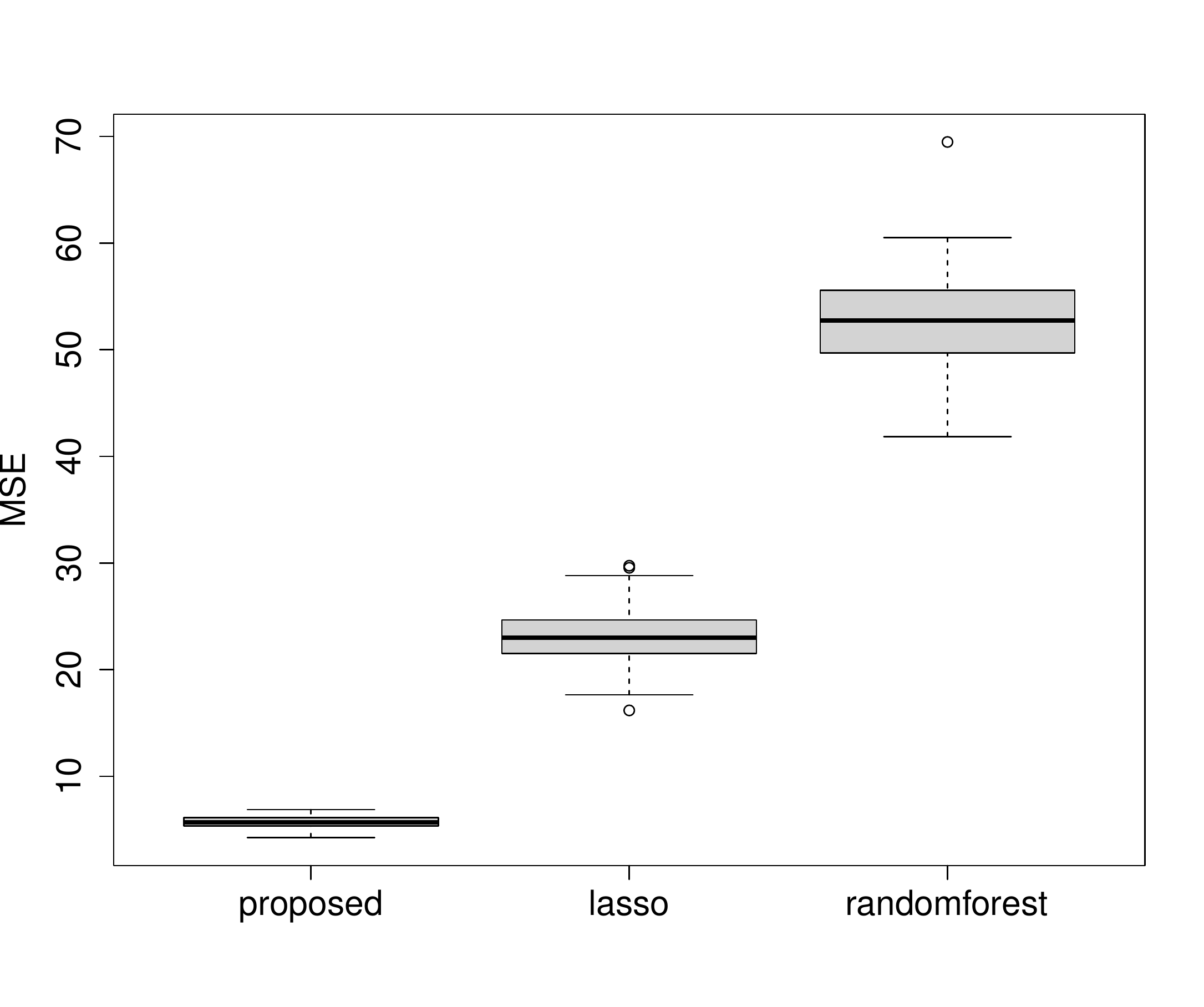}}
\subfigure[Response 1, $\sigma = 1$, and $\rho = 0.8$]
{\includegraphics[width=.45\textwidth,height=.25\columnwidth]{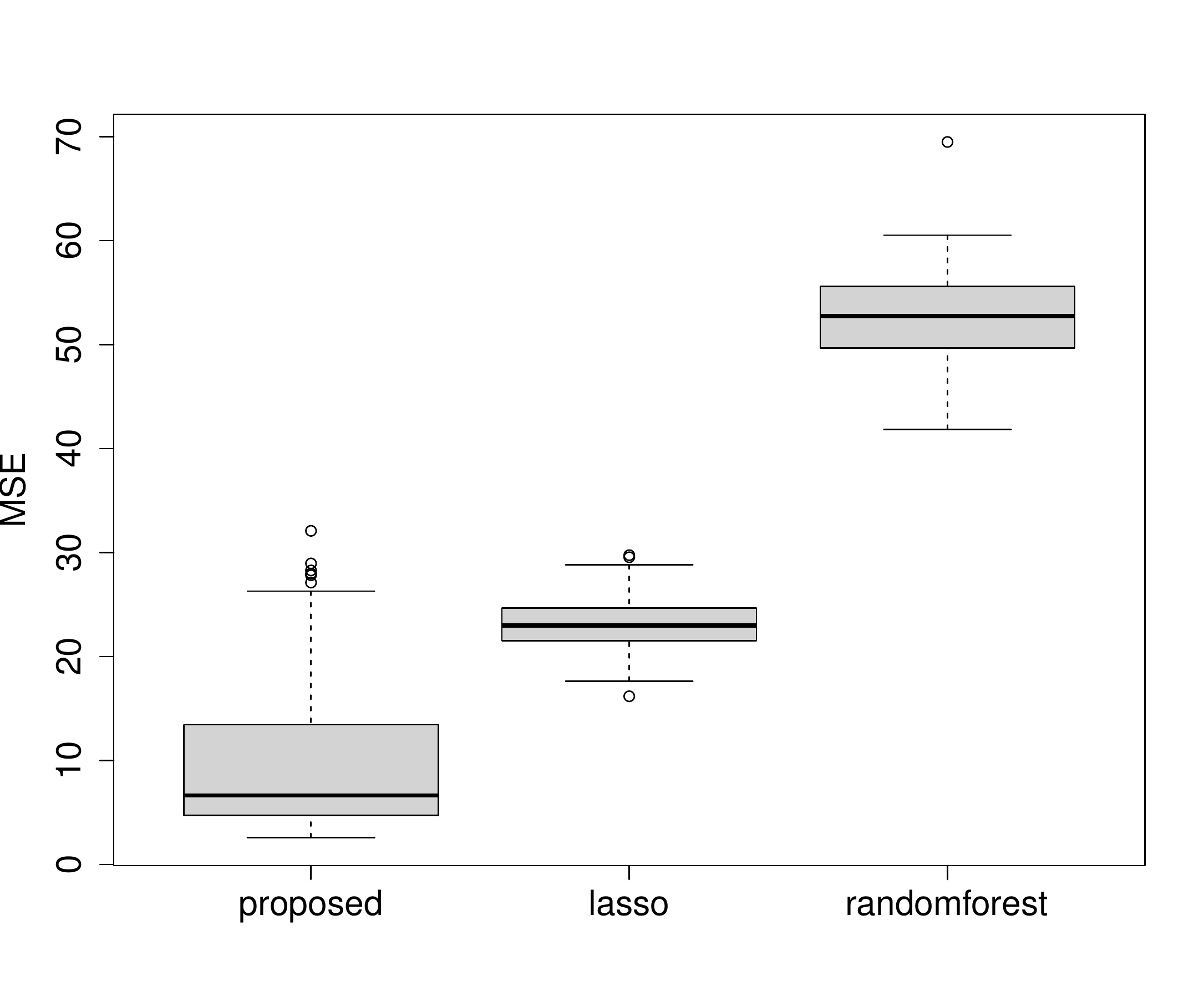}}
\subfigure[Response 1, $\sigma = 2.5$, and $\rho = 0$]
{\includegraphics[width=.45\textwidth,height=.25\columnwidth]{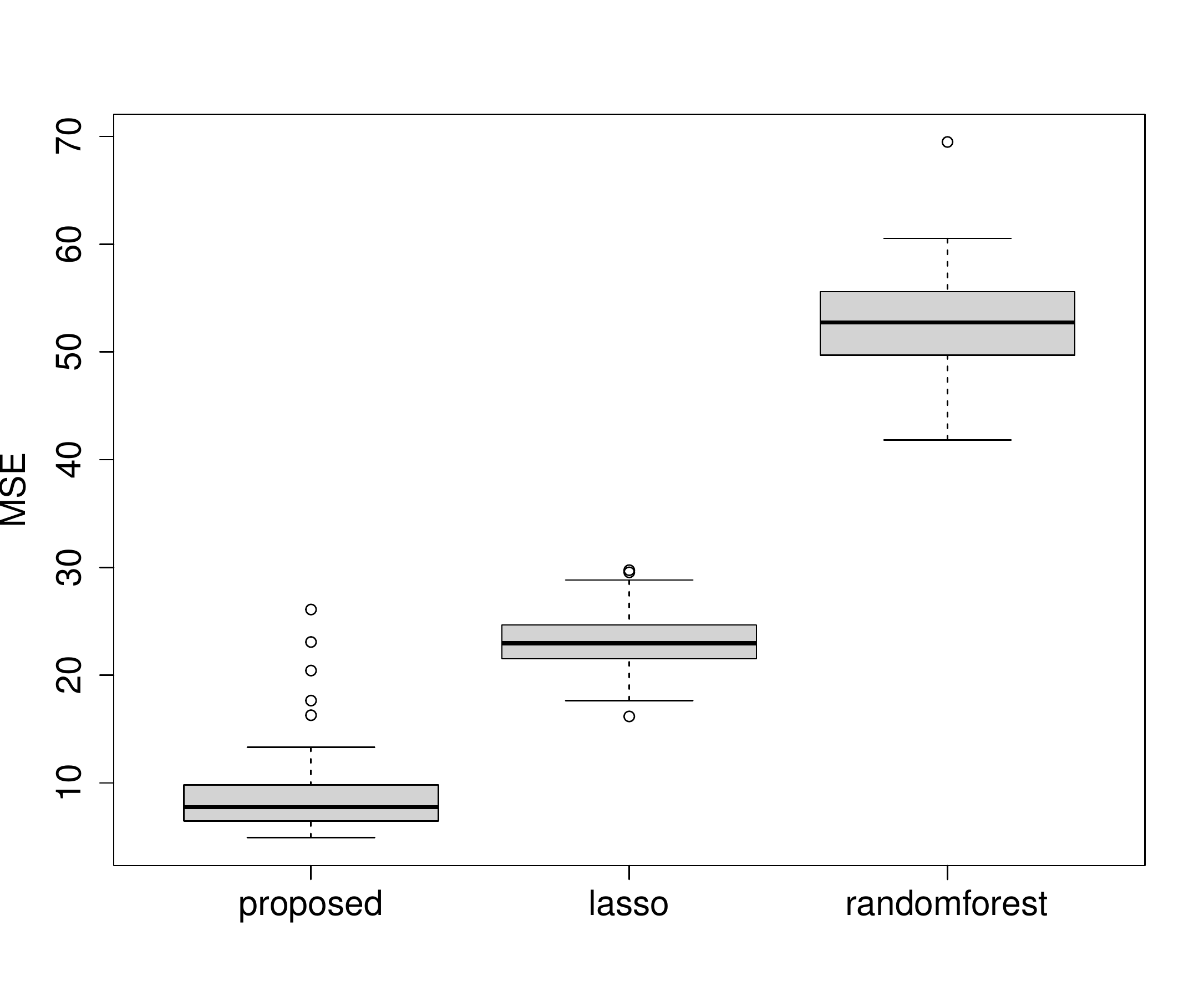}}
\subfigure[Response 1, $\sigma = 2.5$, and $\rho = 0.8$]
{\includegraphics[width=.45\textwidth,height=.25\columnwidth]{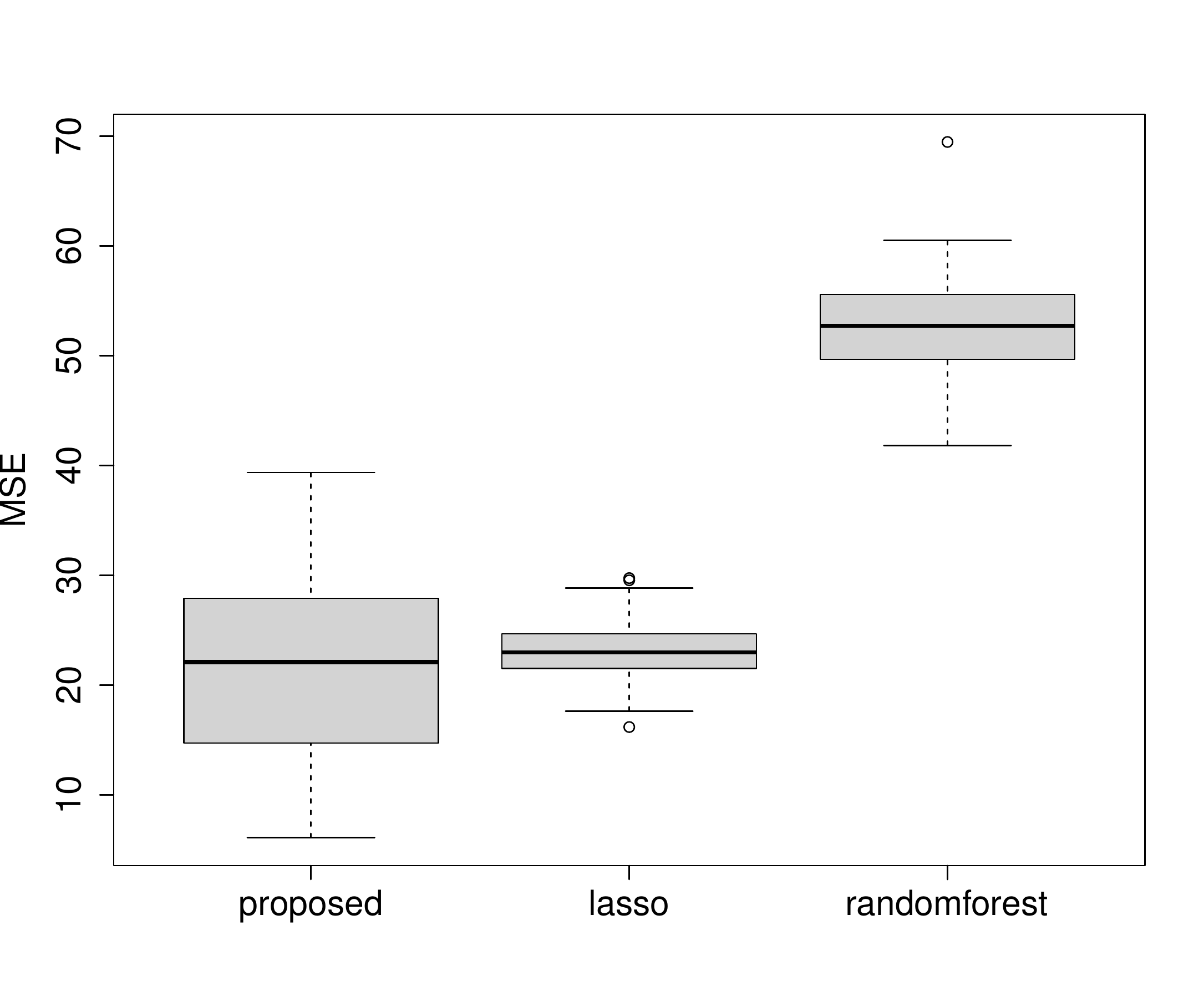}}
\subfigure[Response 2, $\sigma = 1$, and $\rho = 0$]
{\includegraphics[width=.45\textwidth,height=.25\columnwidth]{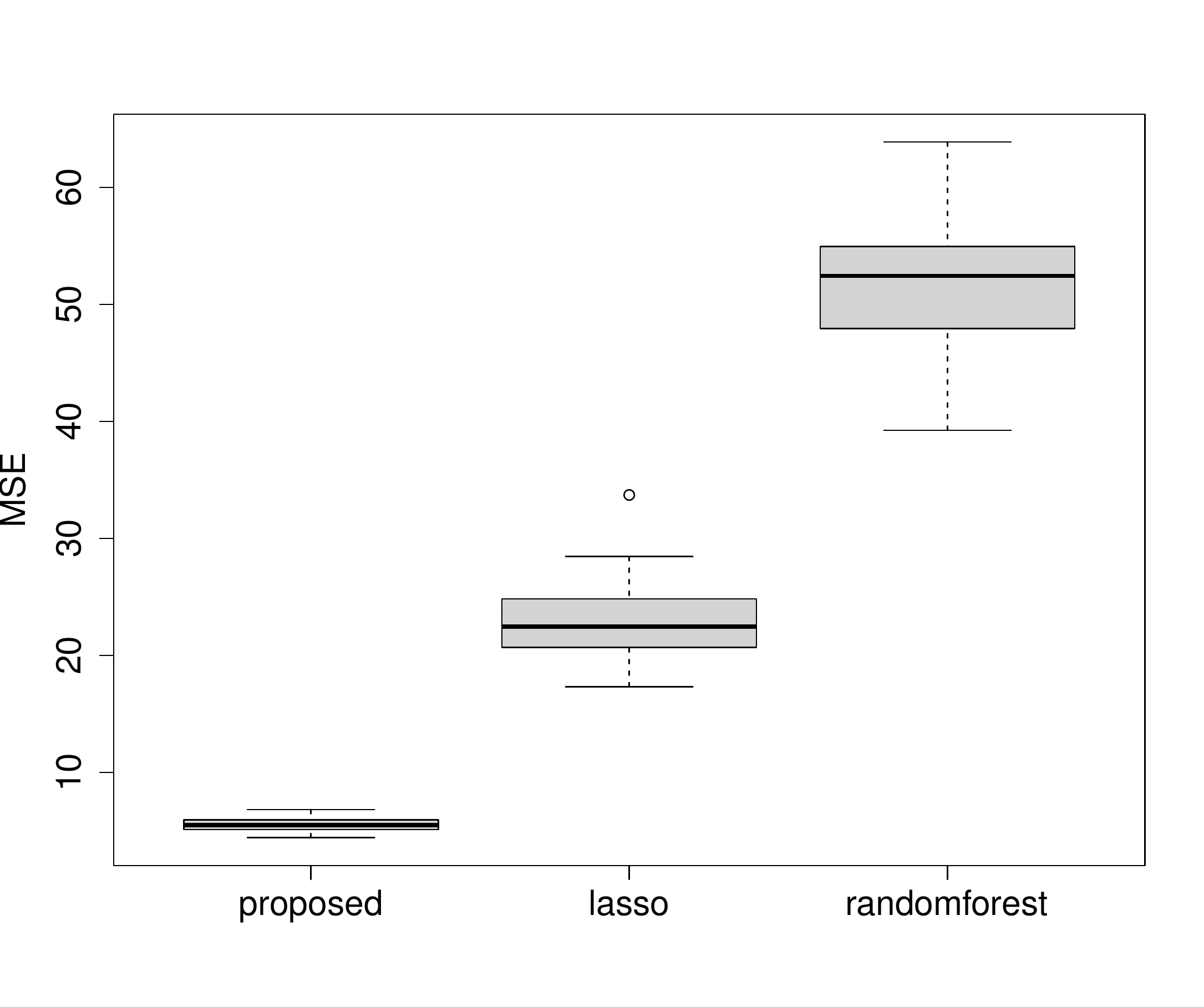}}
\subfigure[Response 2, $\sigma = 1$, and $\rho = 0.8$]
{\includegraphics[width=.45\textwidth,height=.25\columnwidth]{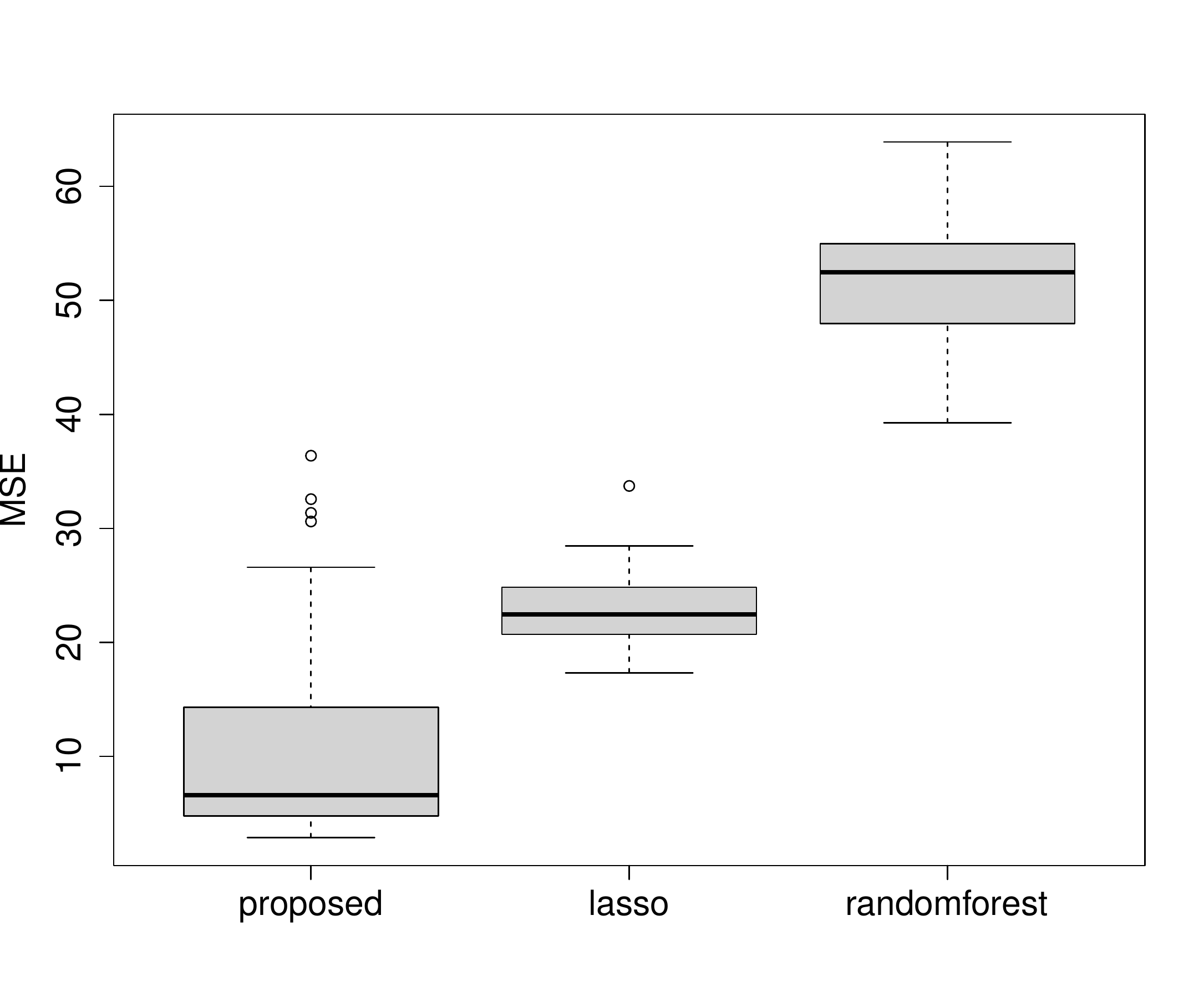}}
\subfigure[Response 2, $\sigma = 2.5$, and $\rho = 0$]
{\includegraphics[width=.45\textwidth,height=.25\columnwidth]{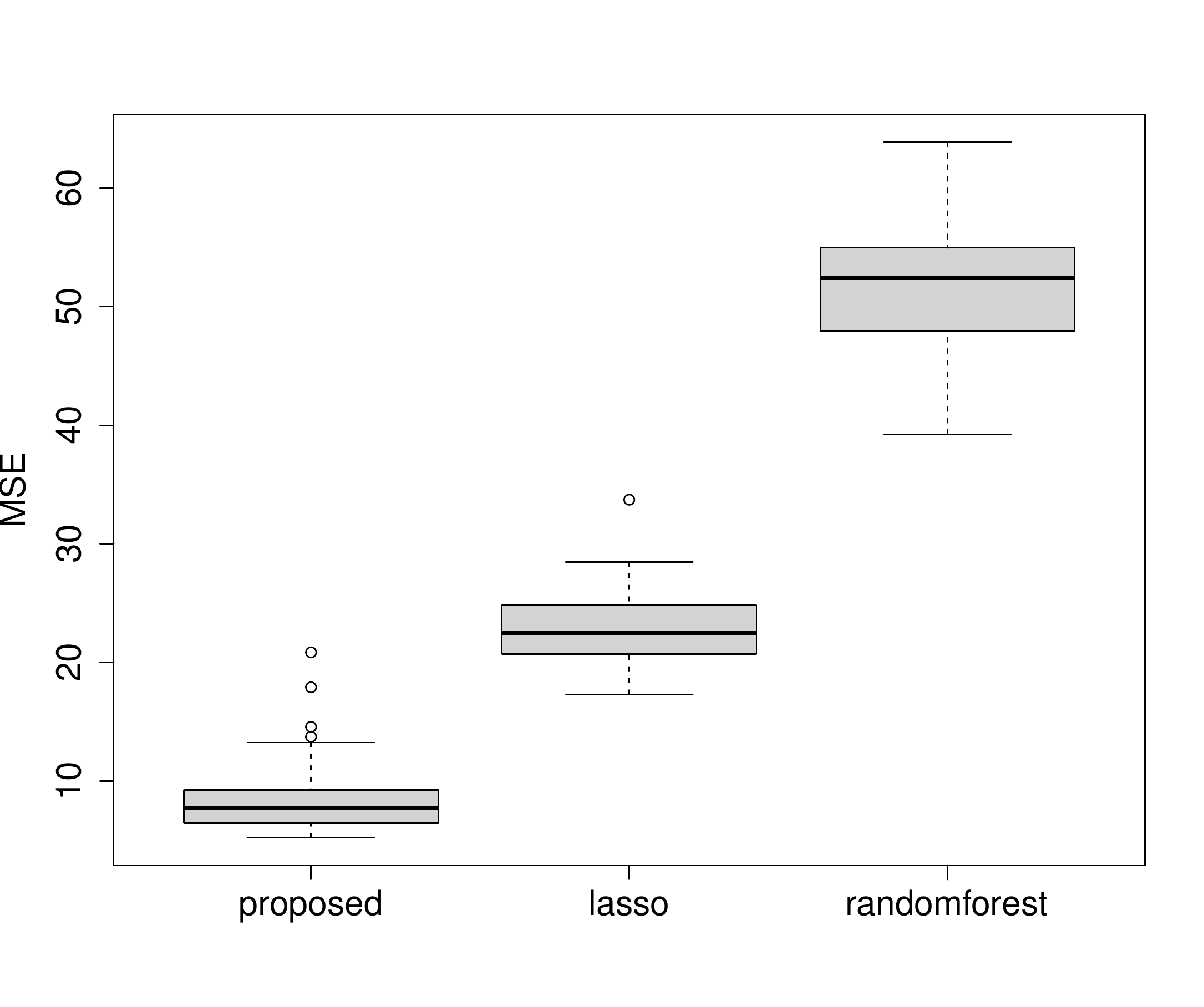}}
\subfigure[Response 2, $\sigma = 2.5$, and $\rho = 0.8$]
{\includegraphics[width=.45\textwidth,height=.25\columnwidth]{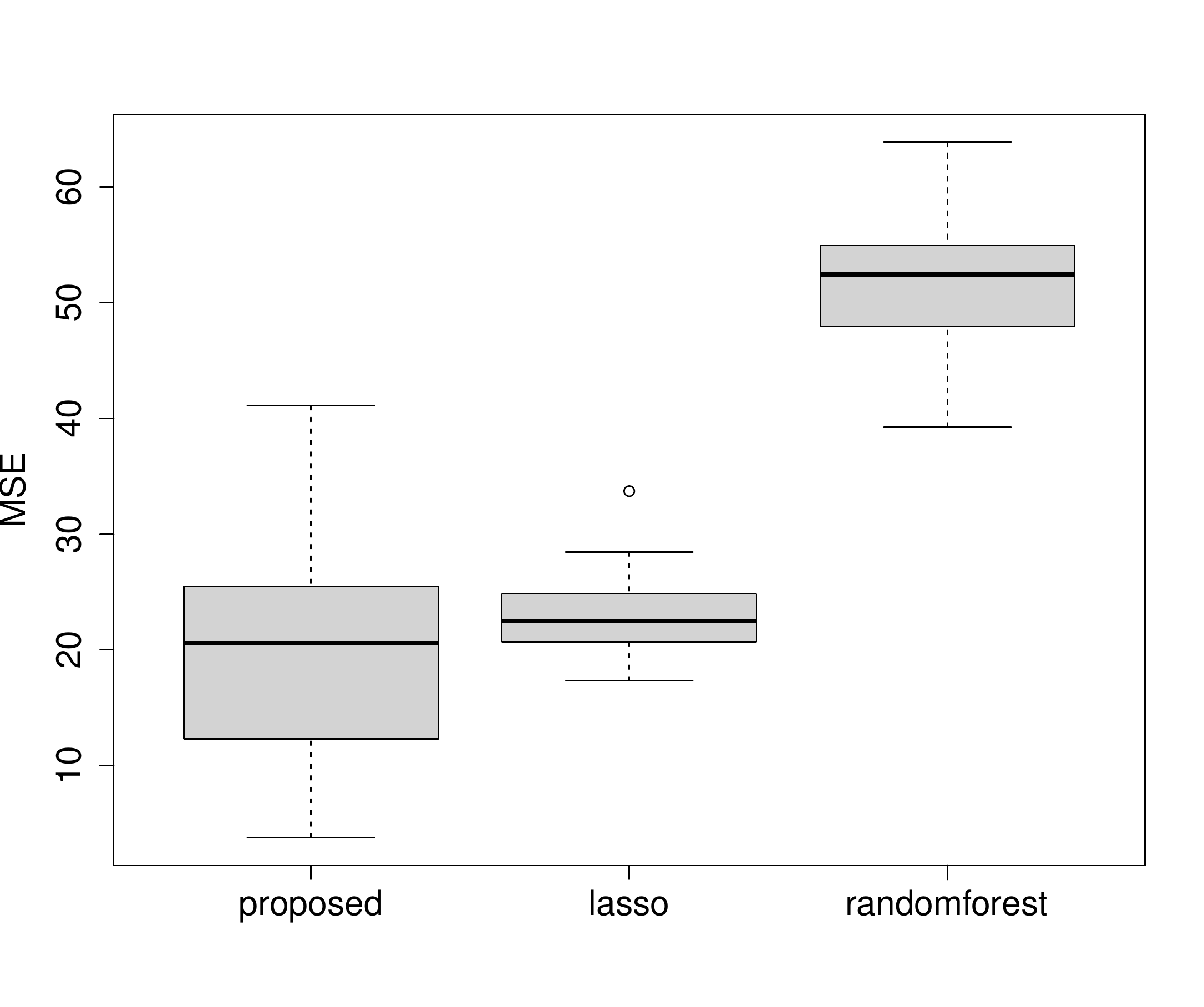}}
\caption{Comparison of the performances of the proposed method, lasso, and random forest under Scenario 2}
\label{fig 2}
\end{figure}

In the second part, we demonstrate the performance of the proposed method compared with the lasso and random forest in the two-response model. For the latter two methods, we treat two models independently. We use MSE as the measure and reveal the results of the three methods under different settings and scenarios in Figures~\ref{fig 1} and \ref{fig 2}. Both figures show that the proposed method generally outperforms others.

\section{Applications}
\subsection{Adult dataset analysis}

In this part, we first analyze an adult dataset to test the effectiveness of the proposed method. The data has been downloaded from the UCI Machine Learning Repository\footnote{https://archive.ics.uci.edu/ml/index.php}. After cleaning the dataset, we have a total of 30160 observations and 11 variables. The response is a $\{1,0\}$ binary variable of individuals who earn higher than \$50,000 or not. We employ nine predictors: age, work class, education num, marital status, occupation, relationship, race, gender, hours per week, and native country. Among them, age, education num, and hours per week are the continuous variables, while the others are categorical variables.

We randomly select 50\% of the total observations for training and the others for testing. Thereafter, we consider the following two situations: (1) we divide the dataset into two parts based on gender and model them separately; and (2) we consider sex as a predictor and model the dataset accordingly. For both situations, we employ the predicted misclassification error and predicted MSE as measures. In the first situation, the predicted misclassification rates for the female and male groups are 0.1157 and 0.3141, respectively, and the predicted MSEs are 0.0752 and 0.1746, respectively. In the second situation, the predicted misclassification rates and the predicted MSEs for the whole dataset modeling are 0.1487 and 0.2466, respectively. When we split the data into two parts based on sex, the predicted error of the female group decreases noticeably, which is only half of the predicted error of modeling the whole dataset.

Further, we compare the proposed method with the lasso and random forest in the two-response model, that is, divide data into two parts following the first situation, as noted above. We obtain the predicted MSEs of the lasso 0.0991 and 0.1921 in the female and male groups, respectively. The former and latter are 132\% and 110\% higher than those of the proposed method, respectively. The predicted misclassification rates of both groups are 0.1153 and 0.3156, respectively. The former is comparable to that of the proposed method and the latter is slightly higher. Meanwhile, we obtain the predicted MSEs of the random forest as 0.1199 and 0.1441 for the female and male groups, respectively, and the predicted misclassification rates are 0.2729 and 0.2138, respectively. The results of the random forest in the male group are lower than those of the proposed method while the results of the random forest in the female group are much higher than those of the proposed method.

\subsection{Right Heart Catheterization dataset analysis}

In this section, we apply the proposed method to a right heart catheterization dataset. The data has been downloaded from the Vanderbilt Biostatistics Datasets\footnote{https://hbiostat.org/data/}. After cleaning the dataset, 36 variables are obtained. The response is a $\{1,0\}$ binary variable of whether a patient died within 30 days after the initial care. It is used to measure the effectiveness of right heart catheterization. Other variables include the records of the diagnoses received by patients, as well as their body dimensions. Similarly, we randomly select 50\% observations for training and the others for testing. We consider the following two situations: first, the dataset is divided into two parts based on the patients’ genders; and second, the entire dataset is modeled and the gender variable is treated as a predictor.

We utilize two measures, namely, the predicted misclassification error and predicted MSE to test the performances of both situations. The predicted MSE and predicted misclassification error of the first situation are better than those of the second. In the first situation, the predicted MSE of the female and male patients are 0.0046 and 0.049, respectively. In the contrast, the predicted MSE of the entire dataset is 0.0052. The predicted misclassification error of the female patients is 0.0024, which is lower than half of that of the entire dataset, 0.0052; and that of the male patients is 0.005, which is also slightly lower than that of the entire dataset.

Further, we compare the proposed method with the lasso and the random forest in the two-response model; specifically, we divide the data into two parts, following the first situation, as noted above. We obtain predicted MSEs of the lasso of 0.0454 and 0.045 in the female and male groups, respectively. The predicted misclassification rates of the two groups are 0.0566 and 0.0551, respectively. Moreover, the predicted MSEs of the random forest are 0.4398 and 0.0067 for the two groups, respectively, and their predicted misclassification rates are 0.5778 and 0.0063, respectively. Most of the errors of the random forest and lasso are much higher than those of the proposed method.

\section*{Acknowledgments}
This work was supported by the National Natural Science Foundation of China (Grant No. 12001557); the Youth Talent Development Support Program (QYP202104), the Emerging Interdisciplinary Project, and the Disciplinary Funding of Central University of Finance and Economics.

\section*{Appendix}
\begin{proof}[Proof of Theorem 1]
To prove Theorem 1 suffices to prove 1) the estimates of the first two steps coincide with the oracle least square estimates; 2) the algorithm converges in the first two steps.
The argument of proving 1) mainly follows arguments of Theorem 5 and Theorem 6 of \citet{tibshirani2021categorical}. The main difference is that we consider the two-response model and \citet{tibshirani2021categorical} considers the univariate response model. In this case, we follow the notations from \citet{tibshirani2021categorical}. We first summarize this fold of proof into five parts. Part 1. Exclude the unpenalized intercept and bound the bound noise item. Part 2. Divide the penalized loss functions according to the predictors. Part 3. Calculate the upper bound between the true coefficients and the oracle least squares estimate. Part 4. Divide the penalized loss functions by ordering the true coefficients. Part 5. Complete the proof and show that $\hat \theta_j = \hat \theta^0_j$ for $j = 1,\dots,p$.
\vspace{.2 cm}

(1) The estimates of the first two steps coincide with the oracle least square estimates. Recall that the ANOVA models relating two responses are
\begin{align}\label{eq model}
& y_1 = \mu_1 + \sum_{j = 1}^p \sum^{K_j}_{k = 1}\theta_{1jk} \mathds{1}_{\{x_j = k\}} + \e_1,\\
& y_2 = \mu_2 + \sum_{j = 1}^p \sum^{K_j}_{k = 1}\theta_{2jk} \mathds{1}_{\{x_j = k\}} + \e_2,
\end{align}
and the first two steps of the proposed algorithm are
\begin{equation}\label{eq solution 11}
\hat \theta_1 =\argmin_{\hat \theta_{1\mathcal{A}_2^c} = 0} \Big\{ \dfrac{1}{2n} \Big\| y_1 - \hat \mu_1 -  \sum_{j = 1}^p \sum^{K_j}_{k = 1}\theta_{1jk} \mathds{1}_{\{x_j = k\}} \Big\|^2_2 + \sum^p_{j=1} \sum^{K_j-1}_{k=1} \rho_{1j} (\theta_{1j(k+1)} - \theta_{1j(k)})\Big\},
\end{equation}
\begin{equation}\label{eq solution 22}
\hat \theta_2 = \argmin_{\hat \theta_{2\mathcal{A}_1^c} = 0} \Big\{ \dfrac{1}{2m} \Big\| y_2 - \hat \mu_2-  \sum_{j = 1}^p \sum^{K_j}_{k = 1} \theta_{2jk} \mathds{1}_{\{x_j = k\}} \Big\|^2_2 + \sum^p_{j=1} \sum^{K_j-1}_{k=1} \rho_{2j} (\theta_{2j(k+1)} - \theta_{2j(k)})\Big\},
\end{equation}
where $\mathcal{A}_2 = \{1,\dots, p\}$ and $\hat \mu_1 = \sum^n_{i=1} y_{1i}/n$; $\mathcal{A}_2$ is obtained from \eqref{eq solution 11} and $\hat \mu_2 = \sum^n_{i=1} y_{2i}/n$.

Part 1. To exclude the unpenalized intercept from the models \eqref{eq model}, for each predictor, we denote the following two vectors:
\begin{align*}
& R^{1j} = \sum^{K_j}_{k = 1} \mathds{1}_{\{x_j = k\}} \hat \theta^0_{1jk} + (I - P^0) \epsilon_1, \\
& R^{2j} = \sum^{K_j}_{k = 1} \mathds{1}_{\{x_j = k\}} \hat \theta^0_{2jk} + (I - P^0) \epsilon_2,
\end{align*}
where $P^0$ is the orthogonal projection onto the linear space, i.e., the residuals from the oracle least squares of both responses are $(I - P^0) y_1 = (I - P^0) \epsilon_1$ and $(I - P^0) y_2 = (I - P^0) \epsilon_2$ respectively. Thus $R^{1j}$ and $R^{2j}$ are the partial residuals of the $j$th predictor respectively.
To control the noise term, we set the following two events
\begin{align*}
& \Lambda_{1jk} = \Big\{ \dfrac{1}{n_{jk}}\big|\mathds{1}^\t_{\{x_j = k\}} (I - P^0)\epsilon_1 \big|  < \dfrac{1}{2} \sqrt{\eta \gamma^*_{j} s_{1j}} \lambda_{1j} \Big\}, \\
& \Lambda_{2jk} = \Big\{ \dfrac{1}{n_{jk}}\big|\mathds{1}^\t_{\{x_j = k\}} (I - P^0)\epsilon_2 \big|  < \dfrac{1}{2} \sqrt{\eta \gamma^*_{j} s_{2j}} \lambda_{2j} \Big\},
\end{align*}
where $j = 1,\dots, p$, and $k = 1,\dots, K$. According to the sub-Gaussian tail bound, we have
\[ P\big((\Lambda^{(1)}_{1jk})^c \big) = P\Big(  \Big\{ \dfrac{1}{n_{1jk}}\big|\mathds{1}^\t_{x_j = k} (I - P^0)\epsilon_1 \big|  \geqslant \dfrac{1}{2} \sqrt{\eta \gamma^*_{j} s_{1j}} \lambda_{1j} \Big\}  \Big)  \leqslant  2 \exp \big( - n_{1jk} \eta \gamma^*_j s_{1j} \lambda^2_{1j}/(8 \sigma^2) \big), \]
where we recall that $\eta \in (0,1]$, $\gamma^*_j = \max\{\gamma_{1j}, \gamma_{2j}, \eta s_{1j}$, and $\eta s_{2j}\}$. Among them, $\gamma_{1j}$, $\gamma_{2j}$ are the tuning parameters of the MCP penalties, $s_{1j}|\{\theta_{1j1},\dots,\theta_{1jK_j}\}| ~\text{and}~ s_{2j} := |\{\theta_{2j1},\dots,\theta_{2jK_j}\}|$.  Similarly we compute $P\big((\Lambda^{(1)}_{2jk})^c \big)$ based on the same tail bound. Then we have, denoting $\Lambda = \cap^{K_j}_{k = 1} (\Lambda^{(1)}_{1jk} \cap \Lambda^{(1)}_{2jk}) $ that
\[P(\Lambda) \geqslant 1 - 2 \exp \big( - n_{j,\min} \eta \gamma^*_j s_{1j} \lambda^2_{1j}/(8 \sigma^2) + \log(2K_j) \big),\]
and in the following, we proceed with the proof based on the event $\Lambda$.

Part 2. Based on the above event, we set $\bar R^{(j)}_{1k} = \mathds{1}^\t_{\{x_j = k\}} R^{(j)}_1$ and $\bar R^{(j)}_{2k} = \mathds{1}^\t_{\{x_j = k\}} R^{(j)}_2$, where $k = 1,\dots, K_j$, and reordering them such that
\begin{align*}
\bar R^{(j)}_{11} \leqslant \cdots  \leqslant \bar R^{(j)}_{1K_j}
~\text{and}~\bar R^{(j)}_{21} \leqslant \cdots \leqslant \bar R^{(j)}_{2K_j}.
\end{align*}
Based on the assumption that $\Delta(\theta_j) \geqslant (4+ 3\sqrt{2}/\eta) (\gamma_j \gamma^*_j)
\max(\lambda_{1j},\lambda_{2j}) $, we have
\begin{align*}
\hat \theta_{1j1} \leqslant \cdots  \leqslant \hat \theta_{1jK_j}
~\text{and}~\hat \theta_{2j1}  \leqslant \cdots \leqslant \hat \theta_{2jK_j}.
\end{align*}
The above results are mainly from Proposition 2 of \citet{tibshirani2021categorical}, which have proved the same results in the univariate optimization. Thus, we omit the proof here. The penalized loss functions for two-response models can be written as follows: for $j = 1,\dots,p$,
\begin{align*}
\hat \theta_{1j} & = \argmin \Big\{\dfrac{1}{2n} \sum^{K_j}_{k = 1} n_{1jk} (\bar R^{(j)}_{1k} - \theta_{jk})^2 + \sum^{K_j - 1}_{k = 1} \rho (\theta_{jk+1} - \theta_{jk}) \Big\},\\
\hat \theta_{2j} & = \argmin \Big\{\dfrac{1}{2n} \sum^{K_j}_{k = 1} n_{2jk} (\bar R^{(j)}_{2k} - \theta_{jk})^2 + \sum^{K_j - 1}_{k = 1} \rho (\theta_{jk+1} - \theta_{jk}) \Big\}.
\end{align*}

Part 3. To calculate the upper bound between the true coefficients and the oracle least squares estimate, we define the following two events,
\begin{align*}
& \Lambda^{(2)}_{1j} = \big\{  \big|\hat \theta^0_{1jk_r} - \theta_{1jk_r} \big| \leqslant \tau_{1j}: l = 1,\dots,s_{1j} | \big\},\\
& \Lambda^{(2)}_{2j} = \big\{  \big|\hat \theta^0_{2jk_r} - \theta_{2jk_r} \big| \leqslant \tau_{2j}: l = 1,\dots,s_{2j} | \big\}.
\end{align*}
Since $y$ is the sub-Gaussian random vector, both $\hat \theta^0_{1jk_l} - \theta_{1jkl}$ and $\hat \theta^0_{2jk_l} - \theta_{2jk_l}$ also follow sub-Gaussian distribution. Based on the sub-Gaussian tail bound and the notations of Theorem~1,
\[ P(\Lambda^{(2)}_{1j})  \geqslant 1 - 2\exp\big(  - c_{\min} \tau^2_{1j}/2\sigma^2 + \log(s_{1j}) \big), \]
and similarly with $\Lambda^{(2)}_{2j}$.

Part 4. Ordering the true coefficients, we obtain that, for $0 =  k_{1j0}<k_{1j1}< \cdots < k_{1js} = K$, $0 = k_{20}<k_{21}< \cdots < k_{2js} = K$ and $j = 1,\dots,p$,
\begin{align*}
& \theta_{1j1} = \theta_{1jk_1} < \theta_{1j(k_1+1)} = \cdots = \theta_{1jk_2} < \cdots < \theta_{1j(k_{s-1}+1)} = \cdots = \theta_{1js},\\
& \theta_{2j1} = \theta_{2jk_1} < \theta_{2j(k_1+1)} = \cdots = \theta_{2jk_2} < \cdots < \theta_{2j(k_{s-1}+1)} = \cdots = \theta_{2js}.
\end{align*}
Based on the results from Part 1 and notations from Part 2, it is directly that $\bar R_{1k_r} - \bar R_{1(k_{r-1}+1)} < \sqrt{\eta \gamma^*_j s_{1j}} \lambda_{1j}$ for $r = 1,\dots,s_{1j} - 1$ and $\bar R_{2k_r} - \bar R_{2(k_{r-1}+1)} < \sqrt{\eta \gamma^*_j s_{2j}} \lambda_{2j}$ for $r = 1,\dots,s_{2j} - 1$. Applying Lemma~\ref{lem 1}, we have
\begin{align*}
&\bar R_{1(k_{r-1}+1)}  \leqslant \theta_{1(k_{r-1}+1)}  \leqslant \hat \theta_{1k_r} \leqslant \bar R_{1k_r} \\
& \bar R_{2(k_{r-1}+1)}  \leqslant \theta_{2(k_{r-1}+1)}  \leqslant \hat \theta_{2k_r} \leqslant \bar R_{2k_r},
\end{align*}
which follow with $\hat \theta_{1(k_r +1)} - \hat \theta_{1k_r} > \gamma_{1j} \lambda_{1j}$ for $r = 1,\dots,s_{1j} - 1$ and $\hat \theta_{2(k_r +1)} - \hat \theta_{2k_l} > \gamma_{2j} \lambda_{2j}$ for $r = 1,\dots,s_{2j} - 1$. Then, we further split the penalized loss functions as follows: for $j = 1,\dots,p$,
\begin{align*}
\hat \theta_{1j} & = \argmin \Big\{  \dfrac{1}{2n} \sum^{K_j}_{k = 1} n_{1jk} (\bar R^{(j)}_{1k} - \theta_{jk})^2 + \sum^s_{j = 1} \sum^{k_r - 1}_{k = k_{r-1} + 1} \rho (\theta_{jk+1} - \theta_{jk})  \Big\} + \dfrac{s_{1j} - 1}{2} \gamma_{1j}\lambda^2_{1j},\\
\hat \theta_{2j} & = \argmin \Big\{  \dfrac{1}{2n} \sum^{K_j}_{k = 1} n_{2jk} (\bar R^{(j)}_{2k} - \theta_{jk})^2 + \sum^s_{j = 1} \sum^{k_l - 1}_{k = k_{r-1} + 1} \rho (\theta_{jk+1} - \theta_{jk})  \Big\} + \dfrac{s_{2j} - 1}{2} \gamma_{2j}\lambda^2_{2j}.
\end{align*}
Subsequently, we obtain the solution separately for each $j$ and $s$.

Part 5. We prove $\hat \theta = \hat \theta_0$ in two parts. First, when $s_{1j} = 1$ or $s_{2j} = 1$, it follows with $n_{1jk} \bar R^{(j)}_{1k}/n = 0$ or $n_{2jk} \bar R^{(j)}_{2k} /n= 0$; then we have $\hat \theta_{1j} = 0$ or $\hat \theta_{2j} = 0$ based on Lemma~\ref{lem 2}.

Now we consider the situation when $s_{1j} > 1$ or $s_{2j} > 1$. We use $j = 1$ and $r = 1$ as an example, and we assume $k_1 > 1$. To obtain the corresponding estimate is equivalent to solving the following functions
\begin{align*}
\hat \theta_{111} & = \hat \theta^0_{111} \mathds{1} + \argmin \Big\{  \dfrac{1}{2n} \sum^{k_1}_{k = 1} n_{1jk} (\widetilde R^{(j)}_{1k} - \theta_{jk})^2 + \sum^{k_1 - 1}_{k = 1} \rho (\theta_{jk+1} - \theta_{jk})  \Big\},\\
\hat \theta_{211} & =\hat \theta^0_{211} \mathds{1}  + \argmin \Big\{  \dfrac{1}{2n} \sum^{k_1}_{k = 1} n_{2jk} (\widetilde R^{(j)}_{2k} - \theta_{jk})^2 + \sum^{k_1 - 1}_{k = 1} \rho (\theta_{jk+1} - \theta_{jk})  \Big\},
\end{align*}
where $\widetilde R^{(1)}_{2k} := \bar R^{(1)}_{2k} -\hat \theta^0_{211}$, $\widetilde R^{(1)}_{1k} := \bar R^{(1)}_{2k} -\hat \theta^0_{111}$, and $\mathds{1}$ is a $k_1$ vector of ones. Note that both $(1/n)\sum^{k_1}_{k = 1}n_{1jk}\widetilde R^{(1)}_{1k} $ and $(1/n)\sum^{k_1}_{k = 1} n_{1jk}\widetilde R^{(1)}_{1k} $ equal zero. Similarly, based on Lemma~\ref{lem 2}, we complete the proof.
\vspace{.2 cm}

(2) The algorithm converges in the first two steps. After the first two steps, we update the nonzero index sets, $\mathcal{A}_1$ and $\mathcal{A}_2$ of $\hat \theta_1$ and $\hat \theta_2$, respectively. We consider the third and the fourth steps and the updated estimates, $\hat \theta_1$ and $\hat \theta_2$. Applying the arguments of (1), we obtain $\hat \theta_1 = \hat \theta^0_1$ and $\hat \theta_2 = \hat \theta^0_1$ and the iteration converges.
\end{proof}

\begin{proof}[Proof of Theorem~2]
Note that, for $l = 1,2$,  with probability at least $1 - 4 \exp \big( - (n_{j,\min} \wedge c_{\min} ) \eta \gamma^*_j s_{1j} \lambda^2_{1j}/(8 \sigma^2) + \log(2K_j) \big)$ that  $$\hat \theta_1  = \hat \theta^0_1 ~\text{and}~\hat \theta_2 = \hat \theta^0_2.$$
By the setting of multiple-response models, for $q$ responses, it is straightforward to calculate that the estimates of the first $q$ steps are equal to the oracle least square estimates of multiple responses, i.e., $\hat \theta_l = \hat \theta^0_l$ for $l = 1,\dots, q$ with probability at least
\[1 - 4 \exp \big( - (n_{j,\min} \wedge c_{\min} ) \eta \gamma^*_j s_{1j} \lambda^2_{1j}/(8 \sigma^2) + \log(q \cdot K_j) \big).\]
Similarly, we obtain that the algorithm converges in the first $q$ steps.
\end{proof}

To prove Theorem~2 and Theorem~2, we require the following two Lemmas, where more details can be found in Lemma 8 and Lemma 10 of \citet{tibshirani2021categorical}.
\begin{lem}\label{lem 1}
Set $l = 1, \dots, q$. Assume $\hat \mu_l = 0$ and $\bar y_{l1} \leqslant \cdots \leqslant \bar y_{lK}$. Further, for $r = 1,\dots, s$ and $j = 1,\dots,p$, assume $\bar y_{lk_r} - \bar y_{l(k_{r-1}+1)} < \sqrt{\eta \gamma^*_{lj} s_{lj}} \lambda_{lj}$ where $k_r$ and $k_{r-1}$ are defined in the proof of Theorem~1. On the other hand, for $r = 1,\dots, s-1$ and $j = 1,\dots,p$, assume $\bar y_{l(k_r +1)} - \bar y_{lk_r} \geqslant \gamma_{lj} \lambda_{lj} + 2(\sqrt{2s_{lj}/\eta} \sqrt{\gamma_{lj}} \lambda_{lj} \vee \gamma_{lj}\lambda_{lj}) + 2 \sqrt{\eta \gamma^*_{lj}s_{lj}} \lambda_{lj}$. Then for $r = 1,\dots, s$, we have $\bar y_{l(k_{r-1}+1)} \leqslant \hat \theta_{l(k_{r-1}+1)} \leqslant \hat \theta_{lk_r} \leqslant \bar y_{lk_r}$.
\end{lem}

\begin{lem}\label{lem 2}
Assume $\sum_k n_{ljk}/n <1$ and $\|\bar y_l\|_\infty < (2 \wedge \sqrt{\gamma_{ljk \sum_k n_{ljk}/n}}) \lambda_{lj} / \sum_k n_{ljk}/n$ for $l = 1,\dots,q$, $j = 1,\dots,p$, and $k = 1,\dots,K$. Then we have $\hat \theta = 0$.
\end{lem}
We omit the proof of the above two lemmas and refer the readers to \citet{tibshirani2021categorical}.

\bibliographystyle{apalike}
\bibliography{C:/reference/reference}
\end{document}